\RequirePackage{etoolbox}
\newbool{fullversion}
\booltrue{fullversion}
\pdfoutput=1
\RequirePackage{etoolbox}

\providebool{fullversion}
\providebool{arxivversion}

\documentclass[acmsmall,screen,nonacm]{acmart}%

\usepackage{preamble}
\usepackage{iris}
\usepackage{prob}
\usepackage{examples}

\ifbool{fullversion}{
  \newcommand{\appref}[1]{\cref{#1}}
  \newcommand{\Appref}[1]{\Cref{#1}}
}{
  \newcommand{\appref}[1]{the Appendix}
  \newcommand{\Appref}[1]{The Appendix}
}

\ifbool{arxivversion}{
\usepackage{caption}
}{
\usepackage[belowskip=-10pt,aboveskip=4pt]{caption}
\setlength{\intextsep}{15pt} %
}

\makeatletter
\ifbool{arxivversion}{}{
\let \MathparLineskip \mpr@lesslineskip %
}
\makeatother

\ifbool{arxivversion}{
  \newcommand{\vsquish}[1]{}
}{
  \newcommand{\vsquish}[1]{\vspace{-#1}}
}

\usepackage{tcolorbox}
\tcbuselibrary{skins,breakable}
  \newtcolorbox{result}{
    blanker,
    extras={interior engine=spartan},
    grow to left by=2pt,left*=0mm,
    grow to right by=2pt,right*=0mm,
    top=1mm,bottom=1mm,
    beforeafter skip balanced=0.1\baselineskip plus 2pt,
    breakable,
    colback=cyan!40
  }

\ifbool{fullversion}{
}{}

\begin{document}

\title{Elton: Urn Resources for Reasoning about Adversarial Probabilistic Programs  %
}

\author[K. H. Li]{Kwing Hei Li}
\orcid{0000-0002-4124-5720}
\affiliation{%
  \institution{Aarhus University}
  \city{Aarhus}
  \country{Denmark}
}
\email{hei.li@cs.au.dk}

\author[A. Aguirre]{Alejandro Aguirre}
\orcid{0000-0001-6746-2734}
\affiliation{%
  \institution{Aarhus University}
  \city{Aarhus}
  \country{Denmark}
}
\email{alejandro@cs.au.dk}

\author[P. G. Haselwarter]{Philipp G. Haselwarter}
\orcid{0000-0003-0198-7751}
\affiliation{%
  \institution{Aarhus University}
  \city{Aarhus}
  \country{Denmark}
}
\email{pgh@cs.au.dk}

\author[J. Tassarotti]{Joseph Tassarotti}
\orcid{0000-0001-5692-3347}
\affiliation{%
  \institution{New York University}
  \city{New York}
  \country{USA}
}
\email{jt4767@nyu.edu}
\authornote{Also affiliated with Amazon Web Services. This paper does not reflect the views of Amazon Web Services.}

\author[L. Birkedal]{Lars Birkedal}
\orcid{0000-0003-1320-0098}
\affiliation{%
  \institution{Aarhus University}
  \city{Aarhus}
  \country{Denmark}
}
\email{birkedal@cs.au.dk}

\begin{abstract}
  Probabilistic programs are important for many applications. For security
  applications in particular, one is  interested in establishing properties
  that hold in the presence of arbitrary adversaries, i.e., unknown pieces of
  code. We present \theaplog, a
  higher-order separation logic for reasoning about higher-order probabilistic programs
  utilizing unknown adversarial code.
  \theaplog incorporates novel logical facilities
  for specifying invariants over distributional properties using \emph{delayed samplings} at the language level,
  and a new kind of separation-logic predicate called \emph{urn resources} at the logic level.
  We show that these extensions are sound and can be erased back to a standard call-by-value semantics.
  Combined with other  features, e.g.~invariants and ghost resources,
  \theaplog is expressive enough to prove error bounds on a wide range of security
   examples, some of which are beyond the scope of previous techniques.
  All proofs are mechanized with the Rocq proof assistant and the Iris separation logic framework.
\end{abstract}

\maketitle

 \section{Introduction}
\label{sec:introduction}

Probabilistic programs are important for many applications, including randomized algorithms,
machine learning, differential privacy, and cryptographic security \cite{miller, morris, murphyml, Dwork:Calibrating:2006, securempc}.
For cryptographic security applications in particular, one is often interested
in establishing properties that hold in the presence of \emph{arbitrary adversaries}
that interact with the cryptographic code in a \emph{security game}.
Classically, adversaries are just modeled as unknown programs (e.g.
probabilistic Turing machines) that interact with the program through some
interface (e.g. a communication channel or some oracle access), and that are
subject to certain restrictions. For example, for studying password security, one is
often interested in so-called preimage resistance of hash functions, i.e., to
prove a bound on the probability that an unknown adversary can
guess a hashed secret (a hashed randomly sampled value), given that the
adversary in the security game can only call the hash function a certain number of
times~\cite{hash-function-basics}.

Most prior work on formal verification of probabilistic security games expresses
the cryptographic program and the game in domain specific modeling languages with special support
for describing adversaries and the interface they use to interact with programs.
However, in the \emph{non}-probabilistic setting, a long line of work~\citep{DBLP:conf/popl/SumiiP04, DBLP:journals/jcs/SumiiP03,robust-safety-gordon,robust-safety}
has instead advocated the use of higher-order languages to model and express security properties.
This higher-order approach has several advantages.
Both the security game and adversaries are expressed as terms in the same language,
with clearly defined semantics that we can reason about using standard language-theoretic techniques.
Adversaries can be modeled as higher-order functions that receive
oracles as a closure, and the language's usual type and abstraction mechanisms can restrict the adversary's access to state of the program.
This is also more trustworthy, as it is comparatively easier to inspect the security game and convince oneself that it
really captures all possible ways an unknown adversary might interact with the code.
This has spawned a long line of work on \emph{robust safety} in the presence of unknown adversarial code.

Given the successes of the higher-order modeling approach for non-probabilistic security properties, it is natural to consider using it for probabilistic properties as well.
However, doing so requires reasoning about the combination of probabilistic choice with higher-order functions and local references, and this is challenging.
In particular, to formally reason about the interactions with unknown code,
one typically needs to define some kind of \emph{invariant} on the state of the system. In
a setting where both the known system and the adversary can perform randomized
actions, the invariant often has to describe a probabilistic state, \ie it is an invariant
about the \emph{distribution} of program values,
and existing verification approaches are often unable to handle expressing these
invariants in the context of a language with higher-order functions and general references.

\begin{figure}
  \centering
\begin{align*}
  \cflip\eqdef{}   \Lam \adv. & \Let \loc = \Alloc (\Flip~\TT) in  \\
                   & \Let (f1, f2) = ((\Lam \_. \loc \gets (\Flip~\TT)), (\Lam \_. \loc \gets \Negb (\deref \loc))) in \\
                   & \adv~f1~f2;  \deref \loc
\end{align*}
\caption{Complicated flip program. }
\label{fig:cflip}
\end{figure}
To illustrate the challenges associated with this class of programs,
consider the program $\cflip$ (short for complicated flip) in \cref{fig:cflip}.
The program $\cflip$ first takes in a higher-order function $\adv$ (an
adversary) as input, which we require to be well-typed with type
$(\tunit \ra \tunit)\ra (\tunit \ra \tunit) \ra \tunit$. The program then
calls $\Flip$, which returns $\True$ or $\False$ with $50\%$ each. It
stores the result in a reference $\loc$ and creates two closures $f1$ and $f2$.
Calling the former overwrites $\loc$ with a freshly-sampled $\Flip$, while
calling the latter reads from the reference and writes back the negated
result to it. The $\cflip$ program then calls $\adv$ with $f1$ and $f2$ passed as arguments.
After $\adv$ returns, $\cflip$ reads from the reference $\loc$.

What can one say about the final read value of $\loc$ in $\cflip$? At
first glance, it might seem like we cannot say anything, since we do not know
anything about the nature of $\adv$. We do not know how many times $\adv$ calls
$f1$ and $f2$, or whether $\adv$ never terminates through unbounded recursion,
or whether $\adv$ uses random sampling internally.

However, the observant reader might notice two things. First, since $\adv$ is
well-typed, it does not have direct access to $\loc$ and the only way it
can modify the contents of $\loc$ is through the invocation of $f1$ and $f2$.
Second, $f1$ and $f2$ respect a probabilistic invariant: if $\loc$ initially
stores a uniformly distributed Boolean, and the decision to call $f1$ and $f2$
is made independently of that value, then after each call to $f1$ and $f2$, the
content is still a uniformly distributed Boolean. Therefore, $\cflip$
eventually returns $\True$ or $\False$ with $50\%$ probability each (modulo the
possibility  $\adv$ never terminates). In other words, $\adv$ can never 
bias the bit stored in $\loc$. Note that this reasoning is subtle, if $\adv$ had read access to $\loc$, this
property would no longer hold; for example, $\adv$ could force $\loc$ to store $\True$
with probability one by reading it, and negating it with $f2$ if it stores $\False$.

Verifying this example formally imposes \emph{three main challenges}: (1) the critical
invariant that captures the probability distribution of $\loc$ is not a property
of a single state, but a \emph{distribution} of states;
(2) the body of $\cflip$ utilizes various hard-to-reason-about programming
features such as \emph{higher-order functions} and \emph{local state}; and (3) the property
holds for \emph{all} possible inputs $\adv$. The above line of reasoning is therefore
\emph{out of scope} of all previous techniques for formal reasoning about
probabilistic programs. There are prior verification techniques that can overcome one or two of
these difficulties, but none of them can tackle all three challenges at
the same time.

To understand why no existing framework can handle these challenges simultaneously,
we can categorize existing probabilistic program logics into two groups based on how they incorporate probabilistic reasoning.
Recall that in most program logics
a predicate describes a set of machine states (possibly with some ghost state).
To account for probabilistic computations,
one approach, which we call  \emph{distributional program logics},
is to instead let predicates describe a set of distributions of machine states.
This is the approach taken by several lines of work~\cite{ellora, psl, lilac, outcomelogic,
  bluebell, amaryllis,   distributional-invariants}. One advantage of this approach is that
it is possible to express directly how a certain program variable is distributed, which
in turn means it is possible to capture how invariants are distributed, e.g.,
the invariant for $\cflip$. However, a disadvantage of this approach is that the semantics
of distributional program logics is more involved, which in turn means that 
existing logics taking this approach lack features needed in order to model the interaction with unknown adversarial code.

In the other approach, which we refer to as \emph{lifting-based program logics}~\cite{HO-UBL, clutch,eris},
one does not change the meaning of predicates, i.e., predicates still describe
sets of machine states. Instead, one changes the meaning of Hoare triples
to account for the fact that a program expression $\expr$ in a Hoare triple
$\hoare{\prop}{\expr}{\propB}$ evaluates to a distribution of machine states.
Since the postcondition $\propB$ still only describes a set of  states,
the semantics of the Hoare triple involves a \emph{lifting} of this predicate to
the whole distribution of  states that $\expr$ evaluates to, hence the
name \emph{lifting-based program logics}. The advantage of lifting-based program
logics is that they inherit all the standard properties of program logics (since
the meaning of predicates does not change), and thus it is easier to extend them
with support for features needed to reason about stateful higher-order functions
and unknown adversarial code. On the other hand, it is trickier to express
distributional invariants, e.g.~the one needed for $\cflip$, since invariant assertions
are only predicates over individual states.

In this work we take a mixed approach. We present Elton, a \emph{lifting-based}
higher-order separation logic for formal reasoning about robust safety of
probabilistic programs in the presence of unknown code. Despite being a
lifting-based logic, Elton supports logical facilities for capturing
distributional properties of the program state, thus recovering some of the
expressivity of assertion-based logics, without giving up on any of the features
needed for higher-order reasoning. We use two main devices to achieve this. First, we
introduce a novel operational semantics with \emph{delayed samplings}, a
distinguished class of suspended probabilistic computations that can be
propagated symbolically and that are only resolved once needed. For example, in
the $\cflip$ program the location $\ell$ will contain such a computation, and
will only be resolved at the end of $\cflip$. We show that
delayed sampling semantics can be soundly erased to standard probabilistic call-by-value semantics.
Second, at the program logic
level, we introduce a novel separation logic resource which we call an \emph{urn
  resource}, which enables us to reason about a \emph{distribution} of states
represented by a delayed sampling, and thereby describe invariants over a
distribution such as that in $\cflip$. \theaplog is inspired by
Eris~\cite{eris}, a separation logic for reasoning about error bounds of
higher-order probabilistic programs with higher-order local state, and \theaplog
inherits all the standard reasoning principles of Eris.\footnote{Currently,
  \theaplog does not support presampling tapes~\cite{clutch,eris}; we
  hypothesize that extending it with tapes is straightforward.} We show that the
combination of urn resources and the reasoning principles of Eris is expressive
enough to reason about probabilistic properties in many adversarial settings,
including the preimage resistance of the random hash oracle and the unlinkability property
of the Basic Hash protocol from the Protocol Ladder
repository~\cite{protocol-ladder}.
In addition, we also prove error bounds for the discrete logarithm
problem~\cite{discrete-log} in the generic group model~\cite{shoup}, which to the best
of our knowledge, is the first formal
proof of this result with respect to a precise operational semantics of program
execution.

\paragraph{Contributions} In summary, we provide
\begin{itemize}
\item The \emph{first} probabilistic higher-order separation logic for reasoning about unknown adversarial code
  that handles the combination of all the three main challenges listed above.
\item A novel operational semantics supporting \emph{delayed samplings} and symbolic execution of
  suspended values.
\item A new separation logic resource called \emph{urn resources} for reasoning about distributions of states arising from delayed samplings.
\item A  soundness proof of \theaplog that ties the operational semantics
	with delayed samplings to standard probabilistic call-by-value semantics.
\item Proofs of a wide range of examples involving adversarial code, including formal accounts of preimage resistance
  of hash functions and the discrete logarithm problem.
\item All results in this paper are mechanized in the Rocq proof assistant~\cite{rocq}, on top of the Iris separation logic framework~\cite{iris, iris2, iris3, irisjournal} and Coquelicot real analysis library~\cite{coquelicot}.
\end{itemize}

\paragraph{Outline}
In the remainder of the paper, we first introduce the key ideas of Elton (\Cref{sec:key-ideas}). Then, after
recalling some preliminaries from probability theory (\Cref{sec:preliminaries}), we 
present the operational semantics of delayed samplings (\Cref{sec:language}) and the Elton program logic (\Cref{sec:logic}).
We then present our case studies (\Cref{sec:case-studies}) before turning to the model and soundness of Elton (\Cref{sec:model}).
We end by reviewing related work (\Cref{sec:related-work}) and conclude with a discussion of limitations and future work (\Cref{sec:conclusion}).%

\section{Key Ideas}
\label{sec:key-ideas}

We give an overview of the main contributions of \theaplog.
We start by recalling the basic rules of error credits from Eris~\cite{eris} in \cref{sec:eris-basics}
and then discuss Eris's limitations in \cref{sec:eris-issue}. We then outline how \theaplog
presents new logical facilities to overcome those issues in \cref{sec:elton-solution}.

\subsection{Basics of Eris}
\label{sec:eris-basics}
\theaplog is an extension of the Eris logic~\cite{eris}.
One of the key features of Eris is the notion of error credits.
We recall the basic rules of error credits
via the following example program:
\begin{align*}
  \guess_1\eqdef{}   \Lam \adv. & \Let i = \adv~\TT in  \\
                              & \Let j = (\Rand \tapebound)+1 in  i=j
\end{align*}
The program $\guess_1$ takes in a well-typed adversary $\adv$ of type
$\tunit \ra \tint$ as  argument, and calls it to return an integer $i$. The program
then generates a secret number $j$ by adding 1 to the result of calling $\Rand \tapebound$,
which in our language returns a number from $\{0,\dots,\tapebound\}$ uniformly
with probability $\frac{1}{\tapebound+1}$ each. Finally,
it returns a Boolean representing whether $i$ is equal to $j$.

Since $\adv$ cannot guess the future random value used to generate $j$,
it is unlikely that $\adv$ returns the value $j$ exactly. In
fact, the probability that the program $\guess_1~\adv$ returns $\True$ is at most
$\frac{1}{\tapebound+1}$, even if $\adv$ utilizes probability internally.
We now explain how we can prove this result  using only the Eris fragment of the \theaplog logic.

At a high level, Eris internalizes the notion of error probabilities with a
separation logic resource called error credits, written as $\upto{\err}$, which
can be spent to exclude undesirable results from random sampling. Proving an Eris
Hoare triple $\vdash \hoare{\prop \sep \upto{\err}}{\expr}{\propB}$ intuitively
means that ``if $\prop$ holds, then the probability $\expr$ returns a result
that violates $\propB$ is at most $\err$''.
The essence of error credits can be boiled down to the following three rules.
\begin{mathpar}
  {
    \infrule[lab]{err-split}{\upto{\err_1 + \err_2}\dashv\vdash\upto{\err_1} \sep \upto{\err_2}}{}
  }
  \and
  \infrule[lab]{err-1}
  {\upto{1}\vdash\FALSE}{}
  \and
    \infrule[right]{ht-rand}
        { \sum_{i=0}^{N}\frac{\Err(i)}{N+1} \leq \err}
        { \hoare{\upto{\err}}{\Rand N}{n . n \in \intrange{0}{N} \sep \Err(n)} }
\end{mathpar}
First, error credits can be split or combined by summation via \ruleref{err-split}. If we hold $\upto{1}$, we can
derive $\FALSE$ with $\ruleref{err-1}$; intuitively, an upper bound of $1$ on an error probability holds trivially.
Last but not least, \ruleref{ht-rand} captures how error credits interact with random sampling. With $\upto{\err}$ in
our precondition, $\Rand\tapebound$ returns a number $n$ between $0$ and $\tapebound$ inclusively,
and we also get $\upto{\Err(n)}$ error credits in the postcondition according to some
error distribution function $\Err$ of our choice, as long as its expected value is upper bounded by $\err$.
As a simple consequence of \ruleref{ht-rand}, one can derive the following \ruleref{ht-rand-avoid} rule. In this rule,
given $\upto{\frac{1}{\tapebound+1}}$ error credits initially, after sampling
from $\Rand \tapebound$, we ensure that the result $n$ is not equal to some integer $m$. 
\begin{mathpar}
  \infrule[lab]{ht-rand-avoid}
        { \hoare{\upto{\frac{1}{\tapebound+1}}}{\Rand N}{n . n\neq m} }{}
\end{mathpar}%
In order to reason about the unknown code $\adv$ in the body of $\guess_1$, we  use a logical relations interpretation of the type system of our language. This gives, for any well-typed program, a specification that  is derived directly from its type, independently from its code.
We do not spell out the details here (see \cref{sec:logical-relations} for more details); for now it suffices to know that
if $\adv$ is well-typed with type $\tunit\ra\tint$, we have  $\vdash\hoare{\TRUE}{\adv~\TT}{v. \exists n\in\integer. v= n}$,
i.e.~$\adv~\TT$ returns some integer.

With these rules at our disposal, we are ready to reason about $\guess_1$.
To prove that $(\guess_1~\adv)$ returns $\True$ with at most probability
$\frac{1}{\tapebound+1}$, it suffices (by the adequacy theorem of Eris, which we
omit here) to show
$\vdash\hoare{\upto{\frac{1}{\tapebound+1}}}{\guess_1~\adv}{v. v=\False}$.
Applying the triple for $\adv~\TT$ that we got from the logical relations interpretation, we deduce that the call to $\adv$ returns some number $i$.
Next, we apply \ruleref{ht-rand-avoid} to
spend the error credits and ensure $\Rand \tapebound$ does not return $i-1$. As a result,
the sum of the sampled value and $1$ is not equal to $i$ and
the expression eventually returns $\False$, satisfying our postcondition.

Let us take a step back and summarize the main ideas of this subsection. Firstly, \theaplog adopts
Eris' approach in using error credits to reason about error probabilities. We can then spend these
error credits to avoid certain undesirable results arising from random sampling. Secondly,
\theaplog provides a logical relations interpretation for reasoning about unknown code; given that some unknown code is
well-typed, the logical relations expresses, in the form of a Hoare triple in the logic, 
that the expression is safe to execute and behaves according to its type.
 
\subsection{Limitations of Eris}
\label{sec:eris-issue}

While using just the features of Eris suffices to handle the $\guess_1$ example, a small change to $\guess_1$ puts it out of scope for Eris.
Consider $\guess_2$, defined below:
\begin{align*}
  \guess_2\eqdef{}   \Lam \adv. & \Let j = (\Rand \tapebound)+1 in \\
                              & \Let i = \adv~\TT in i=j
\end{align*}
The modified program $\guess_2$ is the same as $\guess_1$, except that the order of
the call to the adversary and the random sampling is swapped. To see why
this is a problem, we  attempt to replay our previous
proof, now for $(\guess_{2}~\adv)$, getting us to the following Hoare triple goal:
\begin{mathpar}
  \hoare{\upto{\frac{1}{\tapebound+1}}}{
    \begin{matrix*}[l]
      \Let j = (\Rand \tapebound)+1 in \\
      \Let i = \adv~\TT in  i=j
      \end{matrix*}
}{v. v= \False}
\end{mathpar}
Here, we want to apply \ruleref{ht-rand-avoid} to avoid sampling some value
depending on what $\adv~\TT$ returns, but we have not executed $\adv~\TT$ yet!
Consequently, we have no idea which number $m$ to supply to
\ruleref{ht-rand-avoid}. As a result, to the best of our knowledge, the standard
rules of Eris alone cannot prove the error bound for
$\guess_2~\adv$.\footnote{\citet{completeness-eris} gave a completeness
  proof for Eris, but it only applies to \emph{closed} programs, and so
  does not apply to $(\guess_2~\adv)$.}

Readers familiar with the literature on program logics might wonder
whether this example can be handled using \emph{prophecy variables}~\cite{prophecy1, prophecy2, iris-prophecy},
which allows for logically ``predicting'' what a future value will be during
the course of a proof. Can we not use prophecy
variables to prophesy the value returned by $\adv~\TT$ in advance, allowing us
to know which value $m$ to avoid sampling? Unfortunately, this is not a viable
route to pursue because the standard rules of prophecy variables are \emph{unsound} in the presence of
Eris's other rules for reasoning about random sampling, as we show in \appref{appsec:prophecy}.

Although adding prophecy variables would be unsound, Clutch~\cite{clutch} introduced  a similar technique called \emph{presampling tapes}.
These presampling tapes predict in advance the  outcomes of future $\Rand$ commands.
However, presampling tapes cannot help us predict the value produced by $\adv~\TT$, since we know
nothing about the code of $\adv$ or how it may use randomness to compute its guess.

One approach that \emph{does} work involves a combination of both Clutch and
Eris. First, we can use Clutch and presampling tapes to show that $\guess_1$ and
$\guess_2$ are contextually equivalent, and then the error bound of $\guess_2$
follows from the error bound of $\guess_1$, which we proved in \cref{sec:eris-basics}.
However, this approach is unsatisfactory for two reasons: we need to use \emph{two} different
logics to first rewrite $\guess_2$ and prove a single property, and, as we 
show in later sections, there are examples where this Clutch-plus-Eris approach
also fails (e.g.~\cref{sec:interactive-guessing}).

Another seemingly promising approach to tackle this issue might be to use
a distributional program logic, instead of the lifting-based Eris logic.
For example, by reinterpreting separating
conjunction in separation logic as probabilistic independence \cite{psl},  it might seem
plausible to reason that the random sampling and the call to $\adv~\TT$ are
independent and hence do not interact with each other. However, there are \emph{no}
distributional logics that support rich reasoning principles for
higher-order functions, higher-order references,  and unknown code
(e.g.~$\adv$). One further alternative approach could be to use a denotational semantics and use that to show that
the random sampling and the call to the adversary can be commuted (i.e., to establish the
equivalence of $\guess_{1}$ and $\guess_{2}$), but again there are limitations to such an approach, because
it is unknown how to extend such denotational semantics to randomized programming languages with
general references (higher-order store) as we consider in this paper.

\subsection{Solution with \theaplog}
\label{sec:elton-solution}

Taking a step back, recall that Eris is a lifting-based logic built on top of the Iris base logic, and the possible
truth values of an Eris proposition are the same as for standard Iris separation
logic; there is, a priori, nothing probabilistic about the meaning of a general
Eris proposition --- it is only Hoare triples that talk about probabilistic
behaviours.
As a result, Eris only allows us to
reason about each possible outcome after $\Rand$ independently;
it is not expressive enough
to assert that $j$ is uniformly distributed across $\{1,\dots,\tapebound+1\}$, even after
calling the adversary. 

What we need is a way to make these kinds of distributional assertions \emph{without} giving
up the other expressive features for higher-order programs and general state that Eris has.
\theaplog achieves this through two novel mechanisms.

First, Elton works with a variant of the language extended with a mechanism called \emph{delayed sampling}, a construct that delays
  random sampling actions and propagates their values symbolically throughout execution. 
   For now, we defer most of the details of the changes to the language to \cref{sec:language}.

Second, Elton internalizes delayed sampling via \emph{urn resource} assertions, written $\progurn{\ulbl}{\urn}$.
Intuitively speaking,
$\progurn{\ulbl}{\urn}$ means that the delayed sample variable represented by $\ulbl$ is
uniformly distributed according to the set $\urn$.
  Urn resources allow us to go beyond reasoning about individual
  states and reason about (some) properties of the \emph{distribution} of the program
  state in the logic.
  
With the above extensions, we introduce a bit more probabilistic reasoning
into Elton: instead of having full-fledged assertions about probability
distributions, as in distributional logics like PSL~\cite{psl}, Lilac~\cite{lilac}, and Amaryllis~\cite{amaryllis},
urn resources instead capture how a
\emph{particular} variable is distributed probabilistically.
In many cases this is flexible enough to prove the specifications we want and, moreover,
we do not have to give up the features
needed to build the logical relations model and invariants in our proofs.

To utilize the features of \theaplog to reason about $\guess_2$, we replace all
$\Rand$ constructs with  $\DRand$, a \emph{delayed} variant of $\Rand$ that produces urn resources.
The result is another program $guess_2'$:
\begin{align*}
  \guess_2'\eqdef{}   \Lam \adv. & \Let j = (\hcode{\DRand \tapebound})+1 in \\
                              & \Let i = \adv~\TT in  i=j
\end{align*}
Under the adequacy theorem of \theaplog, presented later in \cref{sec:logic};
proving a Hoare triple about a program with \theaplog implies an upper bound on the
error probability of the same program with the $\DRand$s replaced back with $\Rand$s.
Specifically for this example, by proving a Hoare triple of $\guess_2'$ with \theaplog,
we obtain an error bound for $\guess_2$.
In general, the primitive $\DRand$ is only used for verification purposes and never used
to write any real programs with.

Briefly speaking, the lifecycle of an urn resource $\progurn{\ulbl}{\urn}$ in a proof can be split into three phases: the \emph{Creation}, the \emph{Resolution}, and the \emph{Promotion} phase.
Let us see the urn resource in action by verifying $\guess_2'$, i.e. proving~$\vdash\hoare{\upto{\frac{1}{\tapebound+1}}}{\guess_2'~\adv}{v. v=\False}$.

\begin{figure}[ht]
  \centering
  
\begin{mathpar}
  \infrule[right]{ht-drand}
        { }
        { \hoare{\TRUE}{\DRand \tapebound}{\val . \exists \ulbl. \val=\ulbl \sep \progurn{\ulbl}{\{0,\dots,\tapebound\}}} }
        \and
  \infrule[right]{ht-promote}
  { 
    \hoare{\propC}{\valB}{\propB}
  }
  { \hoare{\rupd{\val}{\propC}{\valB}}{\val}{\propB}}
  \and
  \infrule[right]{ht-resolve-all}
  {
    s\neq\emptyset\\
    \frac{\sum_{i\in s} \Err(i)}{\ssize s}\leq\err\\
     \forall n \in s. \hoare{\upto{\Err(n)}\sep \progurn{\ulbl}{\singleton{n}} \sep \prop}{\expr}{\propB}}
  { \hoare{\upto{\err}\sep \progurn{\ulbl}{s}\sep\prop}{\expr}{\propB}}
\end{mathpar}

  \caption{\label{fig:urn-rules} Selection of urn rules.}
\end{figure}

In the first phase, urn resources are created from the $\DRand$ primitive. Instead of sampling a number uniformly from some distribution, $\DRand$ \emph{creates} an urn, a set representing the possible values a variable can hold, and returns an urn label $\ulbl$, captured by the rule \rref{ht-drand}.
The postcondition of the rule asserts that the urn label $\ulbl$ satifies the urn resouce
$\progurn{\ulbl}{\{0,\dots,\tapebound\}}$, which
states that the urn label $\iota$ represents an integer  distributed uniformly over $\{0,\dots,\tapebound\}$.

After $\DRand$ returns an urn label in $\guess_2'$, we have to add $1$ to it. This brings us to
an important feature of urn labels in the language: we can perform
\emph{computation} on urn labels. Hence, in addition to extending the state of our
language with urns, we also introduce a set of values in our language called
\emph{delayed values}, which are symbolic values representing the result of a (symbolic) computation on
urn labels. For example, in this case, the expression $\ulbl+1$ steps to a
delayed value $\ulbl +_{\delayv} 1$, a special value representing the sum of $\ulbl$
and $1$. We discuss more about these delayed values in \cref{sec:language}.

Once we arrive at the call to $\adv$, 
we apply the logical relation for $\adv$, returning a
natural number $i$. We then perform more computation on the urn labels and
delayed values where the expression $i=(\ulbl +_{\delayv}1)$ steps to the 
delayed value $i=_{\delayv}(\ulbl +_{\delayv}1)$,
bringing us to the following goal state:
\begin{mathpar}
  \hoare{\upto{\frac{1}{\tapebound+1}} \sep \progurn {\ulbl}{\{0,\dots,\tapebound\}}}{ i=_{\delayv} (\ulbl +_{\delayv}1) }
{v. v= \False}
\end{mathpar}
Here we initiate the \emph{resolution} phase of our urn resource via the rule \ruleref{ht-resolve-all} 
in \Cref{fig:urn-rules}.
We can think of this as ``forcing'' the delayed sampling, causing it to draw a random value.
Concretely, given a non-empty urn resource, we can resolve it into any singleton containing an element from the original set,
while distributing the error credits according to a chosen $\Err$ that preserves the expected value. 
This rule is akin to \ruleref{ht-rand} where we can reason about the individual outcomes the variable can take
while distributing error credits in an expectation-preserving manner. As a corollary, we can prove the following
\ruleref{ht-resolve-all-avoid}, the urn-based analogue of \ruleref{ht-rand-avoid}.
\begin{mathpar}
  \infrule[right]{ht-resolve-all-avoid}
  {
     \All n. \hoare{\progurn{\ulbl}{\singleton{n}} \sep n\neq m \sep \prop}{\expr}{\propB}}
  { \hoare{\upto{\frac{1}{\tapebound+1}}\sep \progurn{\ulbl}{\{0,\dots,\tapebound\}}\sep\prop}{\expr}{\propB}}
\end{mathpar}
By applying \ruleref{ht-resolve-all-avoid}, choosing to avoid $m\eqdef{}i-1$, we end up with the goal below:
\begin{mathpar}
  \hoare{\progurn {\ulbl}{\{n\}} \sep n\neq (i-1)}{ i=_{\delayv} (\ulbl +_{\delayv}1) }
{v. v= \False}
\end{mathpar}
By managing to delay resolving the
random sampling until \emph{after} calling the adversary, we are able to spend our error
credits to avoid a value that depended on what the adversary returned. Note however that we cannot establish the
postcondition yet, since our delayed value is not syntactically equal to
$\False$, even though ``semantically'' it represents $\False$ given the urn
resources. Therefore, we have to execute the third phase of our proof, where we
\emph{promote} our delayed value into a proper regular value $\False$ with
\ruleref{ht-promote} from \Cref{fig:urn-rules}. This rule allows us to convert a delayed value
into a simpler one, if it represents the same value semantically under various
urn resources. To do so, it suffices to prove a \emph{promote update predicate}
$\rupd{\val}{\propC}{\valB}$, which intuitively means that the value $\val$ can
be semantically promoted to $\valB$ using the urn resources in proposition $\propC$. We
present more rules of the promote update predicate in \cref{sec:urn-rules}; for now, it
suffices to know that with our current urn resources, we can promote the delayed
value into $\False$:
\begin{mathpar}
  \inferH{}
  {\progurn {\ulbl}{\{n\}} \sep n\neq (i-1)}
  {\rupd{i=_{\delayv} (\ulbl +_{\delayv}1)}{\progurn {\ulbl}{\{n\}}} {\False}}
\end{mathpar}
Hence, by applying \ruleref{ht-promote}, we transform our delayed value back to $\False$, satisfying the postcondition and completing the proof.

To summarize, \theaplog leverages urns to capture distributional characteristics
of variables à la distributional program logics, while retaining the expressive
features from lifting-based logics based on Iris. We \emph{create} urns and allocate their
labels through delayed sampling, and we can perform computation on those labels
to form delayed values which semantically represent actual non-delayed values
in the original program. By \emph{resolving} urns, we reason about smaller subsets of
random outcomes while distributing error credits accordingly. Finally, we
\emph{promote} delayed values back to their normal non-delayed form to match expected
postconditions. By proving a Hoare triple for a program utilizing delayed
sampling, the adequacy theorem of \theaplog implies an error bound for the same
program with the delayed sampling replaced with regular $\Rand$ sampling.

While $\guess_1$ and $\guess_2$ are relatively small synthetic examples, which
can also be verified by other techniques, in later sections (see
\cref{sec:case-studies}), we present considerably larger examples that are
beyond the scope of previous work, especially those involving dynamically
allocated local state and more complex probabilistic reasoning.

\section{Preliminaries}
\label{sec:preliminaries}
This section briefly recalls various concepts in discrete probability theory.
In order to capture the semantics of probabilistic programs that might not terminate, we use \emph{subdistributions}, a generalization of \emph{proper} distributions that admit masses lower than $1$.
  A subdistribution over a countable set $X$ is a function $\distr: X \to [0,1]$ such that $\sum_{x\in X} \distr(x) \le 1$.
  We let $\Distr{X}$ denote the set of all subdistributions over $X$.
For brevity, we simply write \emph{distribution} to mean subdistribution in the rest of the paper and we generally use
the variable $\distr$ to denote a distribution.

For any countable set $X$, the null distribution $\nulldistr: \Distr{X}$ is defined as $\Lam x. 0$.
The uniform distribution $\unifd{\tapebound}$ over $\{0, \dots,\tapebound\}$ is defined as $\Lam n. \text{if}~n\in\{0,\dots,\tapebound\}~\text{then}~\frac{1}{\tapebound+1}~\text{else}~0$.

Subdistributions have monadic operations
\begin{align*}
  &\mret: X\to\Distr{X}\eqdef{}\Lam a~a'. \text{if}~a=a'~\text{then}~1~\text{else}~0 \\
  &\mbind: ((X \to \Distr{Y}) \times \Distr{X}) \to \Distr{Y}\eqdef{} \Lam (f, \distr)~b. \sum_{a \in A} \mu(a) \cdot f(a)(b)
\end{align*}
We write $\mu \mbindi f$ for $\mbind(f, \mu)$.

  The \emph{expected value} of $\Err:X \ra \real$ with respect to distribution $\distr:\Distr{X}$ is defined as $\expect[\distr]{\Err}\eqdef{} \sum_{x\in X}\distr(x)\cdot \Err(x)$. 
  The \emph{support} of  $\mu:\Distr{X}$ is defined as $\supp{\mu} \eqdef{}\{ x \in X \mid 0 < \mu(x)\}$.
  We define the mass of $\distr:\Distr{X}$ as $\mass{\distr}\eqdef{}\expect[\distr]{\Lam x. 1}$.

  The \emph{restriction} of $\distr:\Distr{X}$ to predicate $P$ is defined as ${\restr \distr P}(x)\eqdef{} \text{if}~P(x)~\text{then}~\distr(x)~\text{else}~0$.
Finally, we define the \emph{probability} of a predicate $P$ with respect to $\distr : \Distr{X}$ as $\pr[\distr]{P}\eqdef{}\mass{\restr {\distr}{P}}$.

\section{The \thelang Language}
\label{sec:language}

For concreteness, \theaplog reasons about programs expressed in an ML-like language we call \thelang, though the idea of delayed sampling and urn resources could be transposed to other higher-order languages.
This section defines the syntax~(\cref{sec:language-syntax}) and semantics~(\cref{sec:language-semantics}) of \thelang.
\subsection{Syntax}
\label{sec:language-syntax}

A fragment of \thelang's syntax is given below, with some of the novel primitives \hcode{highlighted}:
\begin{align*}
  \val, \valB \in \Val \bnfdef{}
  & z \in \integer \ALT
  b \in \bool \ALT
  \hcode{\ulbl \in \Ulbl} \ALT
  \TT \ALT
  \loc \in \Loc \ALT
  \Rec \vf \lvar = \expr \ALT
  (\val,\valB) \ALT
  \Inl \val  \ALT
    \Inr \val \ALT\\
  & \hcode{\val +_{\delayv} \valB} \ALT\hcode{\val =_{\delayv} \valB} \ALT\hcode{\Negb_{\delayv} \val} \ALT \ldots
  \\
  \expr \in \Expr \bnfdef{}  &
  \val \ALT
  \lvar \ALT
  \expr_1~\expr_2 \ALT
  \expr_1 + \expr_2 \ALT
  \expr_1 = \expr_2 \ALT
   \Negb \expr \ALT
  \If \expr then \expr_1 \Else \expr_2 \ALT
  (\expr_1,\expr_2) \ALT
  \Fst \expr \ALT \ldots \\
  &\Pack \expr \ALT
   \Unpack \expr as \var in \expr \ALT
  \Alloc~\expr \ALT
  \deref \expr \ALT
  \expr_1 \gets \expr_2 \ALT
  \Rand \expr \ALT
  \hcode{\DRand \expr }
  \\
  \hcode{    s \in \Urn \bnfdef{}}& \hcode{{\mathcal P}_{\text{fin}}(\integer)} \hspace{5em}
  \state \in \State \bnfdef{}  \Loc \fpfn \Val \ \hcode{\times (\Ulbl \fpfn \Urn)}
  \\
  \cfg \in \Conf \bnfdef{}  & \Expr \times \State
\end{align*}
Ignoring the highlighted parts, the syntax is relatively standard, e.g.~we have the usual
first-class recursive function $\Rec \vf \lvar = \expr$ and function application $\expr_1~\expr_2$.
We introduce short-hand notation for lambda abstractions $\Lam x. \expr\eqdef{}\Rec \_ x=e$, let-bindings $\Let x= \expr_1 in \expr_2\eqdef{}(\Lam x. \expr_2)~\expr_1$, and
sequencing $\expr_1; \expr_2\eqdef{}\Let \_ = \expr_1 in \expr_2$.
The language has support for higher-order local state, and we allocate, read from, and write to a reference
with $\Alloc~\expr$, $\deref \expr$, and $\expr_1 \gets \expr_2$, respectively.
Random sampling is performed with $\Rand \expr$.

Looking at the \hcode{highlighted} parts of the grammar, the first addition to the language is the \emph{delayed sampling} construct $\DRand \expr$. As illustrated in \cref{sec:elton-solution},
this is the delayed version of the regular sampling $\Rand \expr$ construct; it is used for verification purposes only,
as the adequacy theorem of \theaplog ultimately asserts properties on regular programs, i.e.~those with all $\DRand$s translated into $\Rand$s.

Delayed sampling returns an \emph{urn label} $\ulbl$.
To represent the results of computation with urn labels, we extend the set of values with \emph{delayed values}, which are boolean and arithmetic expressions over urn labels and constants.
For every possible unary or binary operation, e.g.~$\expr_1+\expr_2$, $\expr_1=\expr_2$, and $\Negb \expr$, the language is extended with a delayed value counterpart, i.e.~$\val+_{\delayv}\valB$, $\val=_{\delayv}\valB$, and $\Negb_{\delayv} \val$, respectively.
Note that from the point of view of the call-by-value semantics, these are still values, so in general they will not be further evaluated.
We also remark that delayed values are syntactic objects only appearing in the evaluation of programs with $\DRand$ and we do not write regular programs with them.

A state $\state$ is represented by a pair of maps. The first component is the heap, a
finite map from locations to values. The second is a
finite map from urn labels to urns.
Each urn is represented by a finite set of integers.
We implicitly project out of the state pair if the projection can be inferred,
e.g.~$\state(\ulbl)$ reads from the urn map if $\ulbl$ is an urn index and
$\lupdate{\state}{\loc}{\val}$ updates the heap if $\loc$ is a heap reference.
A configuration  $\cfg$ is given by a pair of expression and state. 

We equip \thelang with a type system with recursive types, reference types, and polymorphic universal and existential types.
Typing judgments are of the form $\pfctx \mid \vctx \proves \expr : \type$ where $\vctx$ is a context
assigning types to program variables, and $\pfctx$ contains type variables that
may occur in $\vctx$ and $\type$.
The typing rules only apply to the non-delayed fragment of
the syntax, i.e.~the non-highlighted part of the grammar. This means that all
syntactically well-typed programs (particularly in our examples, the adversaries) do \emph{not}
utilize delayed sampling at all internally. Instead, the delayed fragment of the
language only arises in the main program.

\subsection{Semantics}
\label{sec:language-semantics}
\thelang programs follow a call-by-value semantics. 
To explain the full operational semantics of \thelang, we first describe how configurations reduce at each step,
and then explain how to chain those reductions together to capture the full execution of a program.

\paragraph{Single step execution}We define $\stepdistr(\cfg) \in \Distr{\Cfg}$, the function that reduces the configuration $\cfg$ by one step. 
For many cases, the behavior of $\stepdistr$ is unsurprising and we present some examples below, including if-then-else, loading from the store, and random sampling:

\begin{minipage}{.45\textwidth}
  \centering
\begin{align*}
  \stepdistr (\If \True then \expr_1 \Else \expr_2,\state ) &= \mret (\expr_1, \state)\\ 
  \stepdistr (\Rand \tapebound, \sigma) &= \restr{\unifd{\tapebound}\mbindi \Lam n. (n,\state) }{0\leq\tapebound}
\end{align*}
\end{minipage}
\begin{minipage}{.45\textwidth}
  \centering
\begin{align*}
  \stepdistr (\deref \loc,\state ) &=
             \begin{cases}
               \mret(\val, \state) & \state (\loc)=\val \\
               \nulldistr & \text{otherwise}
             \end{cases}
\end{align*}
\end{minipage}
Similar to the allocation of a reference on the heap, the reduction of
$\DRand \tapebound$ allocates an urn in the state, which closely matches the
\emph{creation} rule of urn resources \ruleref{ht-drand}. %
In contrast to reduction of $\Rand \tapebound$, this reduction is 
not probabilistic. It instead deterministically allocates a fresh urn
holding the set $\{0,\dots,\tapebound\}$ representing a number distributed uniformly over this set.
\begin{align*}
  \stepdistr (\DRand \tapebound, \sigma) &= \restr{\mret(\ulbl, \lupdate{\state}{\ulbl}{\{0,\dots,\tapebound\}})}{0\leq \tapebound \land \ulbl= \fresh(\dom(\state))}
\end{align*}
As alluded to earlier in \cref{sec:elton-solution}, urn labels (and delayed values) can be used as
operands for computations, which return delayed values as results. More
specifically, for all unary and binary operators, if all operands are non-delayed
values, we perform the computation as normal. Otherwise, we attempt to evaluate
the expression into a delayed value. This requires type-checking the
operands with a partial function $\typef:\Val \pfn \Type$ to ensure they are of the expected type.
We omit its detailed definition, it just traverses the expression checking the type of every subexpression.
In particular, for  urn label $\ulbl$, we always have $\typef(\ulbl)=\tint$.
Below we show the semantics for addition and negation, for other operators it is defined analogously.
\begin{align*}
  \stepdistr (\val+\valB, \sigma) &=
  \begin{cases}
    \mret(x, \state) & \val, \valB \in \integer \land x=\val+\valB \\
  \mret{(\val+_{\delayv}\valB, \state)}  & \typef (\val)= \tint\land  \typef (\valB) = \tint  \\
   \nulldistr & \text{otherwise}
  \end{cases}\\
  \stepdistr (\Negb \val, \sigma) &=
  \begin{cases}
    \mret(b, \state) & \val \in \bool \land b=\neg \val \\
  \mret{(\Negb_{\delayv} \val, \state)}  & \typef (\val) = \tbool   \\
   \nulldistr & \text{otherwise}
  \end{cases}
\end{align*}
How does the operational semantics capture the \emph{resolution} and
\emph{promotion} phases of urn resources? Perhaps surprisingly,
there are \emph{no} primitives to resolve urns or to promote values with respect
to urn resources. In fact, in many cases, most programs with delayed
sampling get stuck! For example, the expression
$\If (\ulbl =_{v} 0) then \expr_1 \Else \expr_2$ in general gets stuck because
the delayed value in the guard is neither syntactically equal to the boolean
$\True$ nor $\False$.
Instead, resolving urn resources and
promoting values are \emph{ghost operations} only appearing at the logic level
(akin to presampling for tapes in Clutch~\cite{clutch}). It may be helpful to
think of a program with delayed sampling and urns not as a program on
its own, but that as \emph{representing} a distribution of regular non-delayed
programs. 
At the logical level, the rules for resolving urns and promoting
values are sound as they do not affect the \emph{representation} of the
distribution of regular programs, as will be clear when we explain the model of
\theaplog in \cref{sec:model}.

\paragraph{Full program execution}To describe full program execution, we define a stratified execution probability $\exec_{n}\colon \Conf \to \Distr{\Val}$ by induction on $n$:
\begin{align*}
 \exec_{n}(\expr, \state) \eqdef{}
 \begin{cases}
   \nulldistr                                           & \text{if}~\expr \not\in\Val~\text{and}~n = 0, \\
   \mret(\expr)                                         & \text{if}~\expr \in \Val, \\
   \stepdistr(\expr, \state) \mbindi \exec_{(n - 1)} & \text{otherwise.}
 \end{cases}
\end{align*}
Intuitively, $\exec_{n}$ captures execution of a configuration to a value in no more than $n$ steps.
We take the full execution of a configuration $\exec$ as the limit of its stratified approximations, i.e.~$\exec(\cfg)(\val) \eqdef{} \lim_{n \to \infty} \exec_{n}(\cfg)(\val)$, which
is well-defined by the monotonicity and boundedness properties of $\exec_{n}$.
For brevity, we write $\exec \expr$ as notation for $\exec (\expr, \state)$ if it is the same for all states $\state$.

\section{Logic}
\label{sec:logic}

In this section, we give an overview of the \theaplog logic. Since \theaplog is
a higher-order lifting-based logic built on top of the Iris separation logic
framework~\cite{irisjournal}, it inherits many of the propositions and proof
rules from the standard Iris kit, including invariants
$\knowInv{\iname}{\prop}$, ghost resources $\ownGhost{\gname}{\ghostRes}$, and
fancy updates $\pvs[\mask_1][\mask_2]\prop$. We present a collection of those
propositions below:
\begin{align*}
  \prop,\propB \in \iProp \bnfdef{}
  & \TRUE \ALT \FALSE \ALT \prop \land \propB \ALT \prop \lor \propB \ALT \prop \Ra \propB \ALT
  \All \var . \prop \ALT \Exists \var . \prop \ALT \\
  & \prop \sep \propB \ALT \prop \wand \propB %
    \ALT 
    \knowInv{\iname}{\prop} \ALT
    \ownGhost{\gname}{\ghostRes} \ALT
    \pvs[\mask_{1}][\mask_{2}] \prop \ALT \\
    & \upto{\err} \ALT
    \progheap{\loc}{\val} \ALT
    \progurn{\ulbl}{\urn} \ALT
    \rupd{\val}{\prop}{\valB} \ALT
    \hoare{\prop}{e}{\propB}[\mask] \ALT
    \ldots
\end{align*}
Let us focus on the last line of the grammar, which covers the more
\theaplog-specific propositions. Firstly, as mentioned in \cref{sec:eris-basics}, \theaplog
inherits error credits $\upto{\err}$ from Eris to reason about error
bounds.
We use the reference resource $\progheap{\loc}{\val}$ and the urn resource
$\progurn{\ulbl}{\urn}$ to represent exclusive ownership of references and urns,
respectively. \theaplog also introduces a novel proposition called the promote
update predicate, written as $\rupd{\val}{\prop}{\valB}$, which intuitively
expresses that value $\val$ semantically
represents $\valB$ under the assumption of the (urn) resource $\prop$.
This promote update predicate is used mainly for promoting
delayed values, a process that we explain in more detail in \cref{sec:urn-rules}.

Finally, we have the Hoare triple proposition
$\hoare{\prop}{\expr}{\propB}[\mask]$. Here, Hoare triples are sometimes
annotated with the mask $\mask$ to denote the set of invariants that are active
and to prevent re-opening of the same invariants twice; we omit this mask if the
mask is the top mask $\top$. The meaning of Hoare triples is captured 
by the adequacy theorem:

\begin{theorem}[\theaplog Adequacy]\label{thm:adequacy}
  For any $\expr,\expr'\in \Expr$ such that
  $\removedrand (\expr')=\expr$ and
  $\vdash\hoare{\upto \err}{\expr'}{v. \issimpleval~v \sep \pprop~v }$,
  then we have $\pr[\limexecVal~\expr]{\neg \pprop} \leq \err$. %
\end{theorem}
The adequacy theorem is stated in terms of two expressions $\expr$ and $\expr'$,
which are almost identical, except that we obtain $\expr$ by
replacing all $\DRand$ in $\expr'$ with $\Rand$. Given $\upto{\err}$ error
credits in the precondition, by proving a postcondition that states the return value of $\expr'$ 
is a non-delayed value satisfying $\pprop$, we 
can conclude that the
probability that the execution of $\expr$ returns a value \emph{not}
satisfying $\pprop$ is upper-bounded by $\err$. Here, the $\issimpleval~v$ predicate
is satisfied if the value $v$ does not contain any sub-expression or sub-value that
is an urn label or a delayed operation construct (e.g.~the $+_{\delayv}$ construct).
When using this theorem, $\expr$ is the
actual program we want to verify, and we prove its error bound by first rewriting
it to $\expr'$, which utilizes delayed sampling, and then proving a Hoare triple for
$\expr'$, just as we did with $\guess_2$ and $\guess_2'$ in
\cref{sec:elton-solution}.

In addition, we follow \citet{marionneau2026} to
derive a more general adequacy result, which bounds the probability that an expression $\expr$
returns values according to a discrete distribution $\distr$.
\begin{theorem}[\theaplog Distribution Adequacy]\label[corollary]{thm:distribution-adequacy}
  Let $\expr, \expr'\in \Expr$ such that $\removedrand (\expr')=\expr$ and $\mu \in \Distr\Val$.
	If for all non-negative reals $\err$ and bounded non-negative functions $\Err : \Val \to [0,\infty)$ such that $\expect[\distr]{\Err}\leq\err$, we can prove $\vdash\hoare{\upto \err}{\expr'}{v.  \issimpleval~v \sep \upto{\Err(v)}}$, 
  then for all values $\val$, we have $\limexecVal~\expr~\val \leq \distr~\val$. %
\end{theorem}

We divide the rules of \theaplog into three parts. In \cref{sec:logical-relations}, we detail the construction of the logical relations for reasoning about unknown adversarial code. In \cref{sec:basic-rules}, we cover a selection of basic structural and computational rules. Finally in \cref{sec:urn-rules}, we discuss  rules for urn resolution and promotion.

\subsection{Logical Relations}
\label{sec:logical-relations}
As described in \cref{sec:key-ideas}, \theaplog uses \emph{logical relations} to
reason about unknown (well-typed) adversaries. This allows us to derive a Hoare triple
for every well-typed program just from its type, even if we do not know its code.
Specifically, we define a unary
value interpretation $\semInterpS{\type}{\Delta}$, for type $\type$ and semantic
environment $\Delta$, inductively on the type in the same way as in
previous work~\cite{jacm-iris-logrel}. Intuitively, the value interpretation captures values
that behave like valid inhabitants of the type. Since delayed values are just intended as
a proof device, we do not include them in the
value interpretation. For example, even though urn labels semantically represent
(a distribution of) integers, urn labels are not in the value interpretation of
$\tint$. This aligns with the intuition that adversaries should be proper programs,
and should not interact with delayed samples or urn labels.

We then define the semantic typing judgment
$\relates{\Delta}{\expr}{\type}\eqdef \hoare{\TRUE}{\expr}{\val.
  \semInterp{\type}{\Delta}{\val}}$, stating that expression $\expr$ has semantic
  type $\tau$ by asserting that $\expr$ (if
it terminates) evaluates to a value in the value interpretation of $\tau$. We extend the
semantic typing judgment to open terms using closing substitutions, as usual:
$\relates{\Delta\mid\Gamma}{\expr}{\type}\eqdef \forall \gamma.
\semInterp{\Gamma}{\Delta}{\gamma} \wand \relates{\Delta}{\expr\gamma}{\type}$.
where $\gamma$ is a map from the variables in $\Gamma$ to closed values, and $\expr\gamma$
represents the result of substituting those variables in $\expr$ with the values
given by $\gamma$.
By showing that our logical relation is compatible with syntactic typing (via
compatibility lemmas), we can then prove the fundamental lemma of our logical
relations.
\begin{lemma}[Fundamental Lemma of  Logical Relations]\label[lemma]{thm:fundamental}
  If\; \(\pfctx\mid\Gamma \vdash \expr : \tau\)\; then for all $\Delta$ assigning a unary interpretation to all type variables in $\pfctx$, we have \(\relates {\Delta\mid\Gamma} \expr  \tau\)\,.
\end{lemma}
As an immediate corollary, we have the following result for closed programs:
\begin{corollary}[Soundness]\label[corollary]{thm:type-soundness}
  If\; \(\pfctx\mid\cdot \vdash \expr : \tau\)\; then for all $\Delta$ assigning a unary interpretation to all type variables in $\pfctx$, we have $\hoare{\TRUE}{\expr}{\val. \semInterp{\type}{\Delta}{\val}}$\,.
\end{corollary}

\subsection{Basic Rules}
\label{sec:basic-rules}

In this subsection, we cover some of the standard structural and computational rules of \theaplog. We especially focus on those rules that are not directly related to urns and urn resources; we defer the discussion of urn-specific rules to \cref{sec:urn-rules}. We list a collection of basic rules below:%
\begin{mathpar}
  \inferH{ht-bind}
  { \hoare{\prop}{\expr}{\val . \propB}[\mask] \\  \All \val. \hoare{\propB}{\fillctx\lctx[\val]}{\propC}[\mask]}
  { \hoare{\prop}{\fillctx\lctx[\expr]}{\propC}[\mask]}
  \and
  \inferH{ht-frame}
  {\hoare{\prop}{\expr}{\propB}[\mask]}
  { \hoare{\prop\sep\propC}{\expr}{\propB\sep\propC}[\mask]}
  \and
  \inferH{ht-pure}
  {
    \expr \purestep \expr'
    \\ \hoare{\prop}{\expr'}{\propB}[\mask]}{\hoare{\prop}{\expr}{\propB}[\mask]}
  \and
  \inferH{ht-load}
  {}
  {\hoare{l\mapsto v}{\deref l}{w.w=v\sep l\mapsto v}[\mask]}\and
  \inferH{ht-inv-alloc}
  { %
    \hoare{\Exists \iname \in \mask. \knowInv{\iname}{\prop} \sep \propB}{\expr}{\propC}[\mask] }
  { \hoare{\prop \sep \propB}{\expr}{\propC}[\mask] }
  \and
  \inferH{ht-inv-open}
  {\hoare{I\sep\prop}{\expr}{I\sep\propB}[\mask]}
  { \hoare{\knowInv{\iname}{I} \sep \prop}{\expr}{\propB}[\mask\uplus\{\iname\}]}
\end{mathpar}
At a high level, \theaplog inherits many familiar-looking rules from the Iris
logic family. For example, \ruleref{ht-bind} supports modular reasoning by
splitting up a complex expression into a sub-expression and its evaluation
context. Like most, if not all, separation logics, \theaplog comes with the
basic frame rule \ruleref{ht-frame} for framing away some resource $\propC$ from
both the precondition and postcondition. We can take pure steps of an expression
$\expr$ to $\expr'$ with \ruleref{ht-pure}, e.g.~function application,
if-then-else branching, and computation of possibly-delayed values (see
\cref{sec:language-semantics}). We can also symbolically execute the program for heap-manipulating
instructions, e.g.~\ruleref{ht-load} for reading from a reference.

\theaplog comes fully equipped with invariants, which are essential for
building our logical relations (see \cref{sec:logical-relations}) and for reasoning about more
complex examples (see \cref{sec:case-studies}). We can allocate invariants at any point of our
proof via \ruleref{ht-inv-alloc} and we can access the resources of invariants
via \ruleref{ht-inv-open}.
We call attention to the change of masks of the
Hoare triple for \ruleref{ht-inv-open}, where we remove the invariant name from
the mask of the Hoare triple when we open it, thus preventing access to
the invariant again until we close it at the end (by re-establishing the
resources in the postcondition). Unlike in concurrent Iris-based
separation logics, \ruleref{ht-inv-open} can be applied for arbitrary
expressions $\expr$, including those that are not atomic, since \thelang is a sequential language.

\subsection{Urn Rules}
\label{sec:urn-rules}
As mentioned in \cref{sec:elton-solution}, urn resources go through a three-step
\emph{Creation-Resolution-Promotion} routine in a proof. In the first phase, urn
resources are created via the \ruleref{ht-drand} rule (see \cref{fig:urn-rules}). In fact,
this is the only rule that may create urn resources, which makes sense since,
after all, $\DRand$ is the only primitive in \thelang that allocates new urns in
the state (see \cref{sec:language-semantics}). For the rest of this subsection, we dive deeper into
the rules relevant to the other two phases of urn resources.

\paragraph{Resolution of Urns}
Since urns allow us to reason about variables that may take different values, it
is useful to condition on which value an urn takes and reason differently depending
on this value. This is the  motivation behind the
\emph{resolution} of urns. Recall how in $\guess_2'$ in \cref{sec:key-ideas} we
resolved an urn into its individual elements with \ruleref{ht-resolve-all}.
That rule aggressively resolved all of the pending randomness, but
there are various cases
where we want to perform a more fine-grained resolution action where we instead resolve an urn
\emph{partially} to some subset of its possible values.

With this motivation in mind, \theaplog provides \ruleref{ht-resolve-partial},
which can be seen as a generalization of \ruleref{ht-resolve-all}
(i.e. one can derive \ruleref{ht-resolve-all} from \ruleref{ht-resolve-partial}).
Suppose we have an urn $s$.
Instead of immediately resolving it to singleton sets of its members, the user
may provide a partition $l$, that is a list of disjoint subsets of the urn $\urn$,
whose union is well-defined and equal to $\urn$,
describing how the urn should be partially resolved. Then we can resolve
the urn and reason about each case where the urn is resolved to some element in
$l$. Moreover, we can also choose an error distribution function $\Err$, and
distribute error credits for each case, as long as the weighted mean of $\Err$ is
upper-bounded by the initial error credit $\err$ supplied.
\begin{mathpar}
  \inferH{ht-resolve-partial}
  {
    s\neq\emptyset\\
    \emptyset\notin l\\
    \bigcup_{s'\in l} s' = s\\
    \sum_{s'\in l} \Err(s')\cdot \frac{\ssize{s'}}{\ssize s}\leq\err\\
     \forall s' \in l. \hoare{\upto{\Err(s')}\sep \progurn{\ulbl}{s'} \sep \prop}{\expr}{\propB}[\mask]}
  { \hoare{\upto{\err}\sep \progurn{\ulbl}{s}\sep\prop}{\expr}{\propB}[\mask]}
\end{mathpar}
As a corollary, we derive the following rule which allows us to resolve an urn gradually while avoiding one particular value. Unlike \ruleref{ht-resolve-all-avoid}, \ruleref{ht-resolve-partial-avoid} does not resolve our urn to singleton sets, but one where the size is only decreased by $1$. 
Roughly speaking, to prove \ruleref{ht-resolve-partial-avoid}, we apply the rule \ruleref{ht-resolve-partial} where we define $l\eqdef{}[\{m\},\urn\setminus\{m\}]$ and the error distribution function as  $\Err(x)\eqdef{}\text{if}~x=\{m\}~\text{then}~1~\text{else}~\frac{n-1}{\ssize \urn-1}$.
\begin{mathpar}
  \inferH{ht-resolve-partial-avoid}
  {
    m\in s\\
    1\leq n \\
	 \hoare{\upto{\frac{n-1}{\ssize{s}-1}}\sep \progurn{\ulbl}{s\setminus{\{m\}}}\sep \prop}{\expr}{\propB}[\mask]}
  { \hoare{\upto{\frac{n}{\ssize s}}\sep \progurn{\ulbl}{s}\sep\prop}{\expr}{\propB}[\mask]}
\end{mathpar}
  At a high level, $l$ captures the two possibilities of interest. In the first case where the urn is resolved to the singleton $\{m\}$, we also own $\upto{1}$, which can be used to derive $\FALSE$ instantly with \ruleref{err-1}.
  Otherwise, our urn is partially resolved to not contain $m$.
  The error credit re-distribution in this rule allows us to iterate it to remove multiple values. For example, if $n$  is the total number of values we want to avoid in the urn $s$, we can apply \ruleref{ht-resolve-partial-avoid} to avoid a single value first, where we still have $\upto{\frac{n-1}{\ssize{s}-1}}$ for the rest of the proof. Since the size of the new urn $s\setminus\{m\}$ is  $\ssize{s}-1$, these error credits are enough to apply the same rule later on to avoid the other $n-1$ values.
We later provide a wide range of examples in \cref{sec:case-studies} where  \ruleref{ht-resolve-partial-avoid} (and subsequently \ruleref{ht-resolve-partial}) is essential so to support a more gradual resolution pattern of urns.

Recall that urns are sets of integers (see \cref{sec:language-syntax}) and that they represent the distribution where
an integer is sampled uniformly from the set. %
As we later see in \cref{sec:case-studies}, just focusing on representing uniform distributions with urns is sufficient to verify various interesting examples.
If we were to extend our language with other sampling primitives, e.g.~biased coin, we
hypothesize that it is straightforward to extend urns to support arbitrary discrete finite distributions
by representing urns as finite maps. We leave these extensions for future work, as we later discuss in \cref{sec:conclusion}.

\paragraph{Promotion of Values}
In \cref{sec:elton-solution}, we \emph{promoted} a delayed value into a regular value $\False$ by arguing that the delayed value
semantically represents $\False$ under the urn resources via the \ruleref{ht-promote} rule.
To apply the rule, one needs to prove a promote update predicate resource in the precondition;
where roughly speaking, the predicate $\rupd{\val}{\propC}{\valB}$ means that the potentially-delayed value $\val$ semantically represents the non-delayed value $\valB$ under urn resources $\propC$.
With this intuition in mind, let us have a look at a small selection of its rules in \cref{fig:rupd-rules}.

\begin{figure}[ht]
  \centering
\begin{mathpar}
  \inferH{promote-urn}
  {}{\progurn{\ulbl}{\singleton{z}}\vdash \rupd{\ulbl}{\progurn{\ulbl}{\singleton{z}}}{z}}
  \and 
  \inferH{promote-simple}
  {\issimpleval~\val}{\prop \vdash \rupd{\val}{\prop}{\val}}
  \and 
  \inferH{promote-frame}
  { }{(\rupd{\val}{\prop}{\val'}) \sep \propB \vdash \rupd{\val}{\prop\sep\propB}{\val'}}
  \and 
  \inferH{promote-plus}
         { \prop \vdash \rupd{\val}{\prop}{x_1}\\
           \prop \vdash \rupd{\valB}{\prop}{x_2}\\
           x=x_1+x_2
         }
         {\prop \vdash \rupd{(\val+_{\delayv}\valB)}{\prop}{x}}
  \and 
  \inferH{promote-neg}
         { \prop \vdash \rupd{\val}{\prop}{b} \\
           b' = \neg b
         }
         {\prop \vdash \rupd{(\Negb_{\delayv} \val)}{\prop}{b'}}
  \and 
  \inferH{promote-lt}
         { \prop \vdash \rupd{\val}{\prop}{x_1}\\
           \prop \vdash \rupd{\valB}{\prop}{x_2}\\
           b= x_1<_{?} x_2
         }
         {\prop \vdash \rupd{(\val<_{\delayv}\valB)}{\prop}{b}}
  \and
  \inferH{promote-prod}
         { \prop \vdash \rupd{\val}{\prop}{\val'} \\
           \prop \vdash \rupd{\valB}{\prop}{\valB'} 
         }
         {\prop \vdash \rupd{(\val, \valB)}{\prop}{(\val', \valB')}}
\end{mathpar}
  \caption{Selected rules for the promote update predicate. }\label{fig:rupd-rules}
\end{figure}

The first rule \ruleref{promote-urn} captures the notion that urn labels
directly represent the values in the urn they point to. In the case that the urn
contains a single element, the urn label can be directly promoted to that
integer. On the other hand, if a value is not a delayed-value to start with,
e.g.~$\Inr 42$ or $(\True, 67)$, urn resources should not
influence how this value could be promoted as it already semantically
represents itself; hence \ruleref{promote-simple} promotes a non-delayed
value to itself regardless of the resources provided. The
\ruleref{promote-frame} rule allows us to frame away unnecessary urn resources,
e.g.~when resolving the urn label $\ulbl_1$, we can ignore irrelevant urn
resources of other urn labels such as $\progurn{\ulbl_2}{\urn}$.

The promotion of compound delayed values is often syntax-driven, as shown by
the last few rules. The \ruleref{promote-plus} rule states that promoting
$\val+_{\delayv}\valB$ amounts to the sum of two integers resulting from the
promotion of the two operands $\val$ and $\valB$ under the same urn resources
$\prop$. \emph{Mutatis mutandis}, \ruleref{promote-neg} and \ruleref{promote-lt}
promote delayed values under the $\Negb_{\delayv}$ and $<_{\delayv}$ constructors.
This same principle applies for other constructors of
values. For example by \ruleref{promote-prod}, we promote a product by independently
promoting the first and second components.

We remark that there are certain cases where the promotion of delayed values is
valid even when the promotion of its smaller sub-components is ill-formed. Take
the delayed value $\ulbl \times_{\delayv} 0$ or the delayed value
$\ulbl=_{\delayv}\ulbl$, which should morally be promoted to $0$ and $\True$,
respectively, even if $\ulbl$ is pointing to an urn with more than one element.
In these cases, one can use more specific rules involving the
resources in the definition of the promote update predicate (see \cref{sec:weakest-pre}), but
we omit these more specific rules for brevity.

\section{Case Studies}
\label{sec:case-studies}
In this section, we demonstrate the expressiveness of \theaplog on a wide range of complex examples, demonstrating more advanced features of the logic, e.g.~invariants and ghost states. 
\subsection{Interactive Guessing}
\label{sec:interactive-guessing}

We generalize the example $\guess_2$ from~\cref{sec:key-ideas} to the following $\interactiveguess$ program, where now instead of having one attempt to guess the secret $j$, the adversary now has up to $\tapeboundB$ attempts to guess the secret. This example is challenging because the adversary decides on its guesses \emph{interactively} and the number of calls to the adversary is unknown, as we terminate immediately when either the adversary guesses the secret correctly or it runs out of guess attempts. In particular, the Clutch-plus-Eris approach described in \cref{sec:eris-issue} would not work since we cannot rearrange the adversarial calls to be performed first without knowing the exact number of times $\adv$ is called.
\begin{align*}
  \interactiveguess\eqdef{}   \Lam \adv. & \Let j = \ghostdrand \tapebound in \\
                                         & \left(
                                           \begin{array}{l}
                                             \Rec f x = \If x=0 then \False \\
                                             \qquad\qquad \Else \Let i = \adv~\TT in \\
                                             \qquad\qquad \qquad \spac\If i=j then \True \Else f~(x-1)
                                             \end{array}
                                           \right)~\tapeboundB
\end{align*}
Morally speaking, since the adversary cannot access $j$ directly, it has at most $\tapeboundB$ attempts to guess the secret. This means that the probability $j$ is in any one of those $\tapeboundB$ guesses is at most $\frac{\tapeboundB}{\tapebound+1}$. Therefore, we aim to prove the following Hoare triple with error credits $\upto{\frac{\tapeboundB}{\tapebound+1}}$.
\begin{mathpar}
	\hoare{\upto{\dfrac{\tapeboundB}{\tapebound+1}}}{\interactiveguess~\adv}{v. v=\False}
\end{mathpar}
The proof is similar to that of $\guess_2$ in \cref{sec:elton-solution}, except for two main differences. Firstly, with the presence of recursive functions in our program, we apply induction on the number of guess attempts $\tapeboundB$. %
Moreover, the step to resolve urns here is more involved than just applying \ruleref{ht-resolve-all}. Note that we have to  decrease the size of the urn gradually over $\tapeboundB$ steps, so we cannot aggressively resolve the urn to a single value with \ruleref{ht-resolve-all}.
Instead we apply \ruleref{ht-resolve-partial-avoid} repeatedly to partially resolve the urn to avoid each guess attempt from the adversary.

\subsection{Revisiting Complicated Flip}
\label{sec:cflip}

We now revisit $\cflip$ from \cref{sec:introduction} and prove that it returns either $\True$ or $\False$ with $50\%$ probability. Just as we have $\DRand$ for the delayed version of $\Rand$, we implement $\DFlip$ as $\Lam \_.  \DRand 1 = 1$, which can be viewed as the delayed counterpart of $\Flip$.
To show that $\cflip~\adv$ returns $\True$ or $\False$ uniformly, modulo the possibility that $\adv$ might never terminate, we apply \cref{thm:distribution-adequacy}, so that it suffices to prove the following Hoare triple for $\cflip'$, the same program as $\cflip$ but with $\Flip$ replaced with $\DFlip$, for any error $\err$ and bounded non-negative functions $\Err$ such that $\frac{\Err(\True)+\Err(\False)}{2} \leq \err$:
\begin{mathpar}
  \hoare{\upto{\err}}{\cflip'~\adv}{v. \issimpleval~v \sep \upto{\Err(v)}}
\end{mathpar}

To carry out this proof, we first set up some additional definitions to facilitate reasoning.
We can treat $\DFlip$ as its own individual module and provide clean, modular specifications for it. The specification  exposes an abstract value $\flipv$ and an abstract predicate $\flipurn$, which internally is implemented as $\flipv~\ulbl\eqdef{}\ulbl =_{\delayv}1$ and $\flipurn~\ulbl~\urn\eqdef{}\progurn{\ulbl}{(\textlog{set\_map}~\textlog{bool\_to\_Z}~\urn)}$, respectively; they can be seen as the $\DFlip$ counterpart of the regular urn label and urn resource for $\DRand$.

At a high level, the rules of $\DFlip$ (\cref{fig:dflip-rules}) capture the creation-resolution-promotion life-cycle of the $\flipurn$ resource.
The first rule \ruleref{ht-dflip} mirrors that of \ruleref{ht-drand}, when we allocate an abstract urn that stores $\{\True,\False\}$ uniformly. The rule \ruleref{ht-resolve-flip} describes how we can resolve an urn containing $\{\True,\False\}$: we can resolve it to either $\{\True\}$ or $\{\False\}$ and split the error credits between the two cases as long as its average is upper-bounded by the originally-supplied error credits. Lastly, we have \ruleref{promote-flip-urn}, which states that by owning a $\flipurn$ resource that stores a single Boolean $b$, the $\flipv$ value of that urn label can be promoted to $b$.

\begin{figure}[ht]
  \centering
\begin{mathpar}
  \inferH{ht-dflip}
  { }
  { \hoare{\TRUE}{\DFlip~\TT}{\val. \Exists \ulbl. \val=\flipv~\ulbl \sep \flipurn~\ulbl~\{\True,\False\}}[\mask]}
  \and
  \inferH{promote-flip-urn}
  {\flipurn~\ulbl~\singleton{b}}
  {\rupd{\flipv~\ulbl}{\flipurn~\ulbl~\singleton{b}}{b}}
  \and
  \inferH{ht-resolve-flip}
  {
    \frac{\Err(\True)+\Err(\False)}{2}\leq\err\\
     \forall b \in \bool. \hoare{\upto{\Err(b)}\sep \flipurn~\ulbl~\singleton{b} \sep \prop}{\expr}{\propB}[\mask]}
  { \hoare{\upto{\err}\sep \flipurn~\ulbl~\{\True,\False\}\sep\prop}{\expr}{\propB}[\mask]}
\end{mathpar}
  \caption{Specification of $\DFlip$. }
  \label{fig:dflip-rules}
\end{figure}

Next, recall that the program $\cflip$  indirectly shares some reference $\loc$ to the adversary via closures.
Consequently, we need to make use of more complex machinery of \theaplog to define a protocol of how the reference is shared with the adversary, namely \emph{ghost resources} and \emph{invariants}, both of which stem organically from the Iris base logic.

\begin{figure}[ht]
  \centering
\begin{mathpar}
  \infrule[right]{ht-token-alloc}
  {\Exists \gname. \hoare {\ownGhost{\gname}{\token}\sep\prop}{\expr}{\propB}[\mask] }
  { \hoare{\prop}{\expr}{\propB}[\mask]}
  \and
  \infrule[right]{token-excl}
  { }
  { \ownGhost{\gname}{\token}\sep\ownGhost{\gname}{\token}\proves\FALSE }
\end{mathpar}
  \caption{Specification of the exclusive token resource $\token$. }
  \label{fig:token-rules}
\end{figure}

Firstly, we introduce the exclusive token resource $\token$, a standard Iris trick to regulate access to shared resources.
Its two rules are presented in  \cref{fig:token-rules}.  On the left, \ruleref{ht-token-alloc} allows us to allocate a token  with some ghost name $\gname$ at any time of the proof. On the right, \ruleref{token-excl} states that we can derive $\FALSE$ if we have two tokens of the same ghost name in our premise, expressing the idea that each token of a particular ghost name is unique. Intuitively, the token resource is owned by the main program for the full execution of the code, except for the end after calling the adversary, where the main program sacrifices the token to gain ownership of urn resources so we can ultimately resolve the urns for establishing  the postcondition.
Leveraging the token resource, we can now define an invariant $I$ to capture a protocol for sharing a reference $\loc$ with the adversary.
\begin{align}
  I~\gname\eqdef& \Exists (\val : \Val)~(f:\bool\ra\bool).
  \progheap{\loc}{\val} \sep \typef (\val) = \tbool \sep \textlog{bijective}~f \sep \label{eq:cflip-inv-1}\\
          & \quad \left( (\Exists \ulbl.
            \begin{array}{l}
              \flipurn~\ulbl~\{\True, \False\} \sep \\
              (\All b. \flipurn~\ulbl~\singleton{b} \wand \rupd{\val}{\flipurn~\ulbl~\singleton{b}}{(f~b)} )
            \end{array}
            )\lor \ownGhost{\gname}{\token} \right)\label{eq:cflip-inv-2}
\end{align}
Let us go through  the invariant in more detail. In the first line~\eqref{eq:cflip-inv-1}, we include the location resource  so that we can share it with the adversary. We also enforce that the type of the (delayed) value stored in the reference be $\tbool$ according to the partial function $\typef$ (see \cref{sec:language-semantics}), a condition that is used for bookkeeping and ensuring that the delayed value can be used as computation normally according to the operational semantics.
In addition, the invariant states that there exists a bijective function $f$ from Booleans to Booleans. This function $f$ is used to track the relationship between the result produced by a $\DFlip$ and the actual Boolean stored in the reference $\loc$. The other line~\eqref{eq:cflip-inv-2} describes a disjunction. The left disjunct stores a $\flipurn$ containing the set $\{\True,\False\}$ and a resource stating that we can show that the value $\val$ stored in the reference $\loc$ can be resolved to some Boolean $f~b$ when provided with the $\flipurn$ containing the singleton set $\singleton{b}$; the right disjunct only contains the exclusive token resource. Intuitively, the left disjunct captures the general protocol of how the reference is shared with the adversary, while the right disjunct allows the main program to give up the token resource and  obtain the resources in the left disjunct after the adversary program terminates, which is later used with various rules to resolve urn resources for establishing the postcondition. 

With these definitions in place, we can prove the triple about $\cflip$. We start by applying rules to symbolically execute the program, using \ruleref{ht-dflip} at the beginning and allocating a token resource with \ruleref{ht-token-alloc} right before calling the adversary program. We also allocate the invariant with \ruleref{ht-inv-alloc}, instantiating the function $f$ in the invariant to be the identity function, leaving us with the following goal:
\begin{mathpar}
  \hoare{
    \begin{array}{c}
      \knowInv{\iname}{I~\gname}\sep \upto{\err}\sep\\ \ownGhost{\gname}{\token} 
    \end{array}}
  {\begin{array}{l}
    \adv~(\Lam \_. \loc \gets (\Flip~\TT))~(\Lam \_. \loc \gets \Negb (\deref \loc)); \\
    \deref \loc
  \end{array}}{v.
    \begin{array}{c}
      \issimpleval~v \sep \\\upto{\Err(v)}
    \end{array}}
\end{mathpar}
After using the fundamental lemma of the logical relation on $\adv$ to get a triple about it, the remaining goal can be split and simplified into the following three obligations:
\begin{align}
&\hoare{\knowInv{\iname}{I~\gname}}{(\Lam \_. \loc \gets (\Flip~\TT))~\TT}{v. v=\TT} \label{eq:cflip-hoare-1}\\
&\hoare{\knowInv{\iname}{I~\gname}}{(\Lam \_. \loc \gets \Negb (\deref \loc))~\TT}{v. v=\TT} \label{eq:cflip-hoare-2}\\
&\hoare{\knowInv{\iname}{I~\gname}\sep\upto{\err}\sep\ownGhost{\gname}{\token} }
  {\deref \loc}{v. \issimpleval~v \sep \upto{\Err(v)}} \label{eq:cflip-hoare-3}
\end{align}
The first two conditions \eqref{eq:cflip-hoare-1} and \eqref{eq:cflip-hoare-2} arise when we  call the adversary. Recall that $\adv$ has type $(\tunit \ra \tunit)\ra (\tunit \ra \tunit) \ra \tunit$, thus in order to show that $\adv$ can be called safely, we show that the two closures satisfy the value interpretation of $(\tunit \ra \tunit)$ under the invariant $I~\gname$. Proving these two invariants is not too complicated. In both of these goals, we first apply \ruleref{ht-inv-open} to access the resources in the invariant. For \eqref{eq:cflip-hoare-1}, it is analogous to the start of the $\cflip'$ program where we allocate a new $\flipurn$, discarding the old $\flipurn$ in the process. Otherwise for \eqref{eq:cflip-hoare-2}, the result of the reference is negated with respect to the original value, so we instantiate the final function $f$ in the invariant with $\textlog{neg} \circ f'$, where $f'$ is the starting bijective function.

Our last obligation \eqref{eq:cflip-hoare-3} represents the rest of the program execution after calling $\adv$. Here, we apply \ruleref{ht-inv-open} and \ruleref{token-excl} to show that the right disjunct results in a contradiction. With the resources in the left disjunct, we apply \ruleref{ht-resolve-flip} to resolve the $\flipurn$ to hold some singleton set $\{b\}$, splitting our error credits according to $\Err$ and $f$ in the process. Afterwards, we promote the value $\val$ into $f~b$ with \ruleref{ht-promote}. Before we establish the postcondition with $\upto{\Err(f~b)}$, we re-establish our invariant, giving up our token resource to prove the right disjunct of the invariant since we cannot prove the left disjunct with the resolved $\flipurn$ resource.

As mentioned in~\cref{sec:introduction}, if we add a third closure of type $\tunit \to \tbool$ implementing read access, the property no longer holds, that is, the return value might not be uniform. Indeed, our proof would not go through in this case: by the fundamental lemma, our proof would require us to show the following Hoare triple:
\[
	\hoare{\knowInv{\iname}{I~\gname}}{(\Lam \_. \deref \loc)}{v. \exists b\colon \bool. v=b} \\
\]
This is not possible. The postcondition forces us to resolve and promote the urn to a single Boolean value, which means we will not be able to reestablish line~\eqref{eq:cflip-inv-2} in the invariant.

\subsection{Preimage Resistance of Random Oracles}
\label{sec:preimage-resistance}
In cryptography, hash functions are often modeled as random oracles~\cite{uniform-hash-assumption}, where, when the hash of a key $k \in K$ is queried for the fist time, a uniformly chosen result $v \in V$ is sampled and cached, so that subsequent queries of $k$ return $v$. Following \citet{eris}, we implement the random oracle with a mutable map (\cref{fig:hash-code}). When  we query the model with a new key, we sample a random result uniformly, update the map, and return the result. Otherwise, we return the previously stored result. Here we set the set $V\eqdef{}\{0,\dots,\maxsize\}$, i.e.~we sample the result with $\Rand\maxsize$ for some large number $\maxsize$.
\begin{figure}[ht]
  \begin{subfigure}[t]{0.43\textwidth}
\begin{align*}
  \computehash\ m\ v \eqdef{}
   & \MatchML \mapget\ m\ v with
  | \Some(b) => b
  | \None =>
  {\begin{array}[t]{l}
     \Let b = \Rand \maxsize  in \\
     \mapset\ m\ v\ b; b
   \end{array}}
  end {}
\end{align*}
   \end{subfigure}
    \begin{subfigure}[t]{0.4\textwidth}
    \begin{align*}
      \inithash~\_\eqdef{}&\Let m=\initmap~\TT in\\
                          & \computehash~m              
    \end{align*}
 \end{subfigure}
  \caption{Code of random oracle hash model. }
  \label{fig:hash-code}
\end{figure}

There are various basic notions of security for cryptographic hash functions~\cite{hash-function-basics}. In this example, we provide error bounds on the \emph{preimage resistance} of the random oracle hash model under an \emph{interactive} adversary, which we model with a program called $\hashprog$, corresponding to the \texttt{aPre} game in \citet{hash-function-basics}. In this game, after instantiating the hash function, we generate a secret randomly chosen from $\{0,\dots,\tapebound\}$ and query the hash with our secret to obtain $h$. We then create a closure $\hashf'$, which is similar to $\hashf$ except that we write a wrapper around it that limits the number of times the adversary can call $\hashf$ via a mutable counter. We pass both $\hashf'$ and $h$ to the adversary, and we say that the adversary wins this game if it can find some value $g$ such that $h$ is equal to the hash function applied to $g$. 
\begin{align*}
  \hashprog\eqdef{}   \Lam \adv. & \Let (\hashf, \secret) = (\inithash~\TT, \ghostdrand \tapebound) in \\
                               &  \Let h = \hashf~\secret in \\
                               & \Let i=\Alloc \tapeboundB in \\
                               & \Let \hashf'= \left(
                                   \Lam x.   \If \deref i = 0 then \None  \Else (i\gets \deref i - 1 ; \Some (\hashf x))
                                 \right) in\\
                               & \Let g= \adv~\hashf'~h in  \hashf g = h
\end{align*}
We will regard a win by the adversary as an error, and use \theaplog to bound the probability of this error. 
What is a valid upper bound on this probability?
Since the adversary just has to find a value that has the same hash as $h$, they can win in one of two ways: either they guess $h$, or they guess some other value that also has the same hash, \ie they find a collision with the secret value.
Observe that the adversary has at most $\tapeboundB+1$ attempts to probe our hash function.
The first $\tapeboundB$ attempts can be performed during the call to the adversary, while the final attempt comes from hashing the guess $g$ returned by the adversary at the end when we check if the adversary won.
We can thus upper bound the probability of error by summing the probability of the adversary getting the right guess during each probe attempt.
This leads us to prove the following triple:
\begin{mathpar}
  \hoare{\upto{(\tapeboundB+1) \cdot \left(\frac{1}{\tapebound+1}+\frac{1}{\maxsize+1}\right)}}{\hashprog~\adv}{v. v=\False}
\end{mathpar}
where the $1/(\tapebound+1)$ term represents the probability of guessing $h$ during a round, and $1/(\maxsize + 1)$ is the probability of a hash collision.
Since our adversary is \emph{interactive}, we cannot know how the adversary probes our hash function in advance. This suggests that in our proof, we would want to create an urn for our secret via $\DRand\tapebound$, and partially resolve our urn gradually each time the adversary calls our hash function, just as we did previously with $\interactiveguess$ in \cref{sec:interactive-guessing}. A consequence of this is that our urn label is propagated into the hash module, and that we have to devise a good specification for our hash model that not only supports integers as inputs, but \emph{urn labels} or even \emph{delayed values}. This is a significant extension compared to the specifications for hashes found in Eris~\cite{eris} and Coneris~\cite{coneris}.

We present a specification for our hash model supporting delayed values as inputs in \cref{fig:hash-spec}. These rules expose an abstract predicate $\hashfun$ of type $\Val \ra (\Val\pfn\tint)\ra\iProp$ which relates the state of the hash module with an abstract map from (potentially delayed) values to integers. In the first rule \ruleref{ht-init-hash}, we get an empty $\hashfun$ predicate after calling $\inithash~\TT$. The next two rules describe the two cases that occur when applying some hash function $\hashf$ on some key $k$. The rule \ruleref{ht-hash-prev} captures the case where our key has been previously queried by the hash function; in the case that $k$ can be promoted to be equal to some $k'$  in the domain of the abstract map given some urn resources $\propC$ and that it can never be promoted to be equal to anything else in the domain of the map, we deterministically return the value stored by the map under key $k'$. The other rule \ruleref{ht-hash-new} describes the other case where $k$ is new; if $k$ can never be promoted to be equal to anything in the domain of the map under urn resources $\propC$, we update the map with some random value $n$ under the key $k$ and return it. In addition, \ruleref{ht-hash-new} also allows us to distribute error credits according to the return value $n$, \ie by providing some initial error credits $\upto{\err}$ in the precondition, we get back $\upto{\Err(n)}$ as long as  the average of $\Err$ is smaller or equal to $\err$. 

\begin{figure}[ht]
  \centering
\begin{mathpar}
  \inferH{ht-init-hash}
  { \hoare{\TRUE}{\inithash~\TT}{\hashf. \hashfun~\hashf~\emptyset}[\mask]}{}
  \and
  \inferH{ht-hash-prev}
  {\typef (k) = \tint \\
    m(k') = n \\
      \propC \wand 
      (\rupd{k'=_{v}k}{\propC}{\True})\\
  \Sep_{x\in\dom(m)\setminus\singleton{k'}}
    \propC \wand 
    (\rupd{x=_{v}k}{\propC}{\False})
  }
  {\hoare{\hashfun~\hashf~m\sep \propC}{\hashf~k}{n'.n=n' \sep \hashfun~\hashf~m \sep \propC}[\mask]}
  \and
  \inferH{ht-hash-new}
  {\typef (k) = \tint \\
    \sum_{i=0}^\maxsize \frac{\Err(i)}{\maxsize}\leq \err \\
    \Sep_{x\in\dom(m)}
    \propC \wand 
    (\rupd{x=_{v}k}{\propC}{\False})
  }
  { \hoare{\hashfun~\hashf~m\sep\propC\sep \upto{\err}}{\hashf~k}{n. \sep \hashfun~\hashf~(\mapinsert{k}{n}{m}) \sep \propC \sep \upto{\Err(n)}}[\mask]}
\end{mathpar}
  \caption{Specification of hash random oracle. }
  \label{fig:hash-spec}
\end{figure}

The proof of the Hoare triple for $\hashprog$ follows a similar structure as that of $\cflip$ as we are sharing the hash module, a mutable data structure, with the adversary (see \cref{sec:cflip}). At a high level, we symbolically execute the program right until the call of the adversary, where we allocate a token resource and an invariant. After which, we prove that we can call the adversary by showing that the adversary satisfy some value interpretation under the newly-allocated invariant, which involves spending error credits stored in the invariant to resolve urns resources. Finally, we can prove the postcondition by swapping some resources in the invariant with the allocated token resource.
We present details of the invariant and proof sketch in \appref{appsec:preimage-resistance-proof}.

\subsection{Unlinkability of Basic-Hash}
\label{sec:basic-hash}
We use \theaplog to verify a security property of the basic-hash protocol outlined in \citet{protocol-ladder}. In this protocol, we fix a set of valid keys $\{0,\dots,\tagrange\}$.
In order for an individual with key $i$ to authenticate itself to a global reader, it randomly samples a nonce $n$ and sends the pair $(n, hf(n||i))$ to it, where $hf$ is a hash function. The global reader has stored all of the keys, and checks whether the sent hash could be computed from one of the keys it knows of.
In this example, we prove \emph{unlinkability} of the protocol, i.e.~given two sessions of the protocol, it is computationally infeasible for an adversary to distinguish whether the two sessions originate from the same individual.

The unlinkability game for the protocol is modeled in \cref{fig:unlinkability-code}. Like  in \cref{sec:preimage-resistance}, we model the hash function as a random oracle and we begin by initializing a random hash module. This is followed by sampling a uniform random number $b$ from $\{0,1\}$. If $b=1$, both sessions are from the same but randomly chosen individual, otherwise, both individuals are independently sampled. This is evident by the next line where $T_1$ is some random number sampled from $\{0,\dots,\tagrange\}$ and  depending on the value of $b$, $T_2$ is evaluated to either  the same as $T_1$ or some uniformly chosen number. The game then proceeds by sampling the nonces for both sessions, computing the hashes $h_1$ and $h_2$, and creating a closure for probing the hash function, admitting at most $M$ probes. We then pass the tuples $(n_1, h_1)$ and $(n_2, h_2)$, and the closure to $\adv$, after which it returns a guess of the value of $b$. Finally, the adversary wins the game if the guess is equal to $b$.
For simplicity, we only consider two sessions, but it is relatively straightforward to extend the code and proof to arbitrary $n$ sessions.
\begin{figure}[ht]
  \centering
\begin{align*}
  \unlinkgame \eqdef{}   \Lam \adv.&  \Let (\hashf,b) = (\inithash~\TT, \ghostdrand 1) in \\
                                   & \Let T_1 = \ghostdrand \tagrange in \\
                                   & \Let T_2 = b \cdot T_1 + (1-b) \cdot (\ghostdrand \tagrange) in \\
                                   & \Let (n_1, n_2) =(\Rand \noncerange, \Rand \noncerange) in \\
                                   & \Let (h_1, h_2) = (\hashf~(n_1 \cdot (\tagrange+1) + T_1), \hashf~(n_2 \cdot (\tagrange+1) + T_2)) in \\
                                   & \Let i=\Alloc \tapeboundB in \\
                                   & \Let \hashf'= \left(
                                       \Lam x.   \If \deref c = 0 then \None \Else (c\gets \deref c - 1 ; \Some (\hashf x))
                                     \right) in\\
                                   & \Let g= \adv~(n_1, h_1)~(n_2, h_2)~\hashf' in g=b
\end{align*}
  \caption{Code of basic-hash unlinkability game. \label{fig:unlinkability-code} }
\end{figure}

We prove the following Hoare triple with \theaplog, which asserts that the adversary has advantage proportional to $\mathcal{O}(\frac{1}{\noncerange+1} + \frac{\tapeboundB}{\tagrange+1})$. 
\begin{mathpar}
  \hoare{\upto{\frac{1}{2}+ \frac{1}{\noncerange+1} + \frac{2\cdot \tapeboundB}{\tagrange+1}}}{\unlinkgame~\adv}{v. v=\False}
\end{mathpar}
Looking at the precondition of the Hoare triple, we see that the error credit provided is the summation of three components.  Firstly, the $\upto{\frac{1}{2}}$ error accounts for the $\frac{1}{2}$ chance that the adversary blindly guesses the $b$ correctly. The $\upto{\frac{1}{\noncerange+1}}$ error is used to eliminate the small probability that the two nonces for both sessions coincide; if they are the same, the adversary can guess with high probability whether the two sessions come from the same individual by directly comparing both hashes. Lastly, the last component of the error credit is used to resolve the urns representing the identity of the individuals, ensuring all inputs to the closure $\hashf'$ do not coincide with their identifiers. Since the adversary has at most $\tapeboundB$ attempts to probe the hash, and we have two urns to resolve, this sums up to $\upto{\frac{2\cdot \tapeboundB}{\tagrange+1}}$ in the precondition.

The key idea in the proof is similar to that of \cref{sec:preimage-resistance}, where error credits are spent via \ruleref{ht-resolve-partial-avoid} such that the set of integers represented by $h_1$ and $h_2$ does not contain the input provided to the hash function $\hashf'$ by the adversary. For reasons of space, we refer readers to \appref{appsec:basic-hash-proof} for the invariant definition.

\subsection{Generic Group Model}
\label{sec:generic-group}

The security of cryptographic protocols often relies on the hardness of various problems, e.g.~the discrete logarithm problem~\cite{discrete-log}, which states that given a cyclic group of prime order $p$, generator $g$, and some arbitrary group element $g^x$, it is difficult to compute $x$. Just as we use a random oracle to model hash functions, we use the \emph{generic group model}~\cite{shoup}, an idealized cryptographic model, to analyze the upper bound of computational power necessary to break a hardness assumption of an arbitrary cyclic group. The generic group model can be seen as an oracle that generalizes the notion of a cyclic group. Working at this level of abstractions lets us study classes of attacks that do not rely on exploiting the concrete group representation (a.k.a.~\emph{generic algorithms}).

\begin{figure}[ht]
  \centering
\begin{align*}
  \dlog\eqdef{}   \Lam \adv. &  \Let (x,\varr) = (\ghostdrand (p-1), \arrnew~\TT) in \\
                               & \Let (g1, gx)=(\arrpush~\varr~1, \arrpush~\varr~x) in \\
                              & \Let f = \vtryspend~(\Alloc \tapeboundB) in \\
                              & \Let (\vmulf, \vinvf, \veqf) = (\vmul~\varr~f, \vinv~\varr~f, \veq~\varr) in \\
                              & \Let \vall =\Pack (g1, gx, \vmulf, \vinvf, \veqf) in \\
                               & \Let x' = \adv~\vall in x' = x
\end{align*}
  \caption{Code of discrete logarithm game. \label{fig:discrete-log-code} }
\end{figure}
The group model encodes each group element as a unique identifier; a generic algorithm can then utilize the oracle to perform specific group operations by querying it with  identifiers, and receiving the identifier corresponding to the result.
In this example, we prove an error bound for an arbitrary generic algorithm to break the discrete logarithm problem; we present a snippet of the code in \cref{fig:discrete-log-code} and more details, e.g.~the implementation of the generic group, can be found in \appref{appsec:generic-group-proof}. At a high level, after taking an adversary $\adv$ as input, $\dlog$ first randomly samples $x$ uniformly from $\{0, \dots, p-1\}$ for prime number $p$, and initializes our group model with $\arrnew$. We then generate two identifiers $g1$ and $gx$ from the group model, which represents the generator $g$ and group element $g^x$. Subsequently, we create $\vmulf$, $\vinvf$, and $\veqf$, which are functions for $\adv$ to perform multiplication, inverse, and equality on identifiers, respectively. We bound the number of times $\adv$ can perform multiplication and inverse operations via the $f$ closure (similar to \cref{sec:preimage-resistance}). We then pack $g1$, $gx$, and the group operations as an existentially typed value to be passed to $\adv$, returning a guess $x'$. The adversary wins the discrete logarithm game if its guess matches $x$.
Our error bound proven  is equal to the error bound of  $\mathcal{O}(\tapeboundB^2/p)$ proved by \citet{shoup}, where $\tapeboundB$ is the number of operations $\adv$ can perform.
Specifically, we prove the following Hoare triple:
\begin{mathpar}
  \hoare{\upto{\frac{2+\tapeboundB\cdot(\tapeboundB+3)/2}{p}}}{\dlog~\adv}{v. v=\False}
\end{mathpar}
Compared to previous examples, this case study is significantly more involved; we defer details of our implementation, invariant, and proof sketch to \appref{appsec:generic-group-proof}.
Here we remark that this example is challenging for various reasons. Firstly, compared to a previous machine-checked proof of \citet{formalization-generic-group}, our model of the adversary is \emph{interactive}, i.e.~it can decide on the future group operations according to the results of previous ones. Secondly, our adversary is typed with the existential type, to ensure that the only way it can interact with group identifiers is through specific group operations, thus highlighting higher-order features of \theaplog. Thirdly, we have to utilize another kind of ghost resources to define the invariant and the abstract type of these group identifiers, namely $\authm(\nat_\text{max})$.
Lastly, unlike previous examples like \cref{sec:interactive-guessing} and \cref{sec:preimage-resistance}, error credits are spent to eliminate not one, but  multiple values in an urn for each group operation the adversary performs.
To the best of our knowledge, we are the first to formally mechanize the error bounds for the discrete logarithm problem \wrt interactive generic algorithms.

\section{Model}
\label{sec:model}
We now define the semantic model of \theaplog. In \cref{sec:weakest-pre}, we present the  weakest precondition modality, which gives meaning to Hoare triples. Subsequently in \cref{sec:soundness}, we give an overview of the proof of the adequacy theorem, \cref{thm:adequacy} from \cref{sec:logic}.
\subsection{Weakest Precondition}
\label{sec:weakest-pre}
Following \citet{irisjournal}, Hoare triples in \theaplog are expressed in terms of a weakest precondition, i.e.~$\hoare{\prop}{\expr}{\propB}[\mask] \eqdef{} \prop \wand \wpre{\expr}[\mask]{\propB}$. This weakest precondition is defined in two steps. Firstly, we define an intermediate $\wpreinner{\expr}{\pred}$ predicate, expressed as a guarded fixpoint where all recursive occurrences exist after a later modality $\later$. We then define $\wpre{\expr}[\mask]{\pred}$ by wrapping the start and end of the fixpoint with fancy updates. This structure is slightly different from other probabilistic lifting-based logics~\cite{eris, coneris} where the weakest precondition is defined as one fixpoint; this change enables us to open invariants across non-atomic steps (see \ruleref{ht-inv-open} from \cref{sec:basic-rules}).
\begin{align*}
  \wpreinner{\expr_{1}}{\pred} \eqdef{}
  & \All \state_{1}, \err_{1} .
    \stateinterp~\state_{1}~\err_{1} \wand \stateStepl{\expr_1}{\state_{1}}{\err_{1}}\spac\{\expr_2, \state_{2},  \err_{2} \ldotp \\
  & \quad \big(\expr_{2} \in \Val \sep
    \stateinterp~\state_{2}~\err_{2} \sep \pred~\expr_{2}\big) \lor{} \\
  & \quad
    \big( \expr_{2} \not\in \Val \sep
    \progStepl{\expr_{2}} {\state_{2}}{\err_{2}}
    \spac\{ \expr_{3}, \state_{3}, \err_{3} \ldotp  \\
  & \qquad \later \stateStepl{\expr_{3}}{\state_{3}}{\err_{3}}
    \spac\{ \expr_4, \state_{4}, \err_{4} \ldotp \stateinterp~\state_{4}~\err_{4} \sep \wpreinner{\expr_{4}}{\pred} 
    \} \} \big) \}\\
  \wpre{\expr}[\mask]{\pred}\eqdef{}& \pvs[\mask][\emptyset]\wpreinner{\expr}{v. \pvs[\emptyset][\mask]\pred~v}
\end{align*}
The definition of the intermediate predicate $\wpreinner{\expr_1}{\pred}$ starts with a separating implication that allows a prover to assume ownership of the \emph{state interpretation} $\stateinterp~\state_{1}~\err_{1}$, which gives meaning to heap resources $\progheap{\loc}{\val}$, urn resources $\progurn{\ulbl}{\urn}$, and error credits $\upto{\err}$. This definition is standard and we refer readers to previous Iris-based logics~\cite{irisjournal, eris}.
The definition of  $\wpreinner{\expr_1}{\pred}$ utilizes a state step $\stateStepp$ and a program step precondition $\progStepp$; we provide their definitions later. Intuitively, $\stateStepp$ describes urn-specific actions such as urn resolution and value promotion, while $\progStepp$ captures individual steps of program execution.

 We structure these preconditions into a $\stateStepp$-$\progStepp$-$\stateStepp$ chain, similar to the weakest preconditions of Approxis~\cite{approxis} and Coneris~\cite{coneris}. At a high level, we initially prove a $\stateStepl{\expr_1}{\state_1}{\err_1}$ precondition, resulting in a new expression $\expr_2$, state $\state_2$, and error budget $\err_2$. This is followed by a case split on whether $\expr_2$ is a value. If it is, we return the updated state interpretation and prove that $\expr_2$ satisfies postcondition $\pred$. In the other case, we execute one program step with $\progStepp$, and prove another $\stateStepp$ precondition. At the end, we reestablish the state interpretation and prove $\wpreinner{\expr_4}{\pred}$, where $\expr_4$ is the final resulting expression.

Before we focus on the details of $\stateStepp$ and $\progStepp$, we present two auxiliary definitions.
Firstly, we define the function $\urnsf: \State\ra\Distr{\Ulbl\fpfn\tint}$ that takes in a state as input and returns a distribution of substitution functions, represented by partial functions from urn labels to integers. This $\urnsf$ function captures the intuition of picking a single element from each urn uniformly. We define $\urnsf$ inductively on the urn map of the input. If it is empty, we return the partial function with empty domain with probability $1$. Otherwise, we randomly sample one element uniformly from the first urn, and add that to the result returned by $\urnsf$ in the recursive call.
\begin{align*}
  \urnsf~m \eqdef{}&
                            \begin{cases}
                              \mret~(\Lam x. \textlog{None})& \text{urns}(m)=\emptyset\\
                              \unifd{s}\mbindi(\Lam x. \urnsf~m' \mbindi (\Lam f. \mret \lupdate{f}{\ulbl}{x})) & m = \lupdate{m'}{\ulbl}{s} \land \ulbl\notin \dom (m')
                            \end{cases}
\end{align*}
Given a substitution function $f: \Ulbl\fpfn\tint$, we have functions $\urnsubst$ and
$\urnsubstcfg$ performing urn substitutions on values and configurations,
respectively. (They also replace all $\DRand$ expressions with $\Rand$ as well.)
We do not give their precise definitions here, but intuitively they are
straightforward syntax-directed functions that replace delayed values with
proper non-delayed values according to $f$, returning a regular value or
configuration without $\DRand$ or delayed values.

Secondly, we define the $\erasable$ predicate. We say a distribution of states $\distr$ is erasable with respect to state $\state$ if the distribution of substitution functions computed by $\distr$ followed by $\urnsf$ is the same as $\urnsf~\state$, i.e.~$\erasable(\distr, \state)\eqdef{} \distr \mbindi \urnsf = \urnsf~\state$.
As we later see, we bake the definition of erasable distributions into the state step precondition, supporting urn resolution.
\paragraph{State Step Precondition}
The state step precondition is defined as an inductive fixpoint of four inference rules (\cref{fig:stateStep}).
We have two base cases:  \ruleref{state-step-err-1} states that if the error budget argument is larger or equal to $1$, we can prove any $\stateStepp$ precondition.  \ruleref{state-step-ret} states that the precondition holds if the arguments satisfy the postcondition argument $\pred$.

As mentioned before, $\stateStepp$ is defined to support urn-related actions, as captured by the next two rules. The rule \ruleref{state-step-exp} allows us to perform any erasable action $\distr$ \wrt the state argument while distributing errors across the branches of $\distr$ in an expectation-preserving manner. This rule is used to prove the validity of urn resolution; in particular, we prove that the resolution of urns via \ruleref{ht-resolve-partial} (see \cref{sec:urn-rules}) is an erasable action.
Lastly, the last inference rule \ruleref{state-step-promote}  captures urn promotion (see \ruleref{ht-promote} in \cref{sec:elton-solution}). If we can substitute the value $\val$ into $\valB$ by urn substitution under any function $f$ under the support of $\urnsf~\state$, we can promote $\lctx[\val]$ to $\lctx[\valB]$, where $\lctx$ is some evaluation context.

\begin{figure*}[htb]
  \centering
  \begin{mathpar}
    \infrule[right]{state-step-err-1}
    {1\leq\err}
    { \stateStep{\expr}{\state}{\err}{\Phi} }
    \and
    \infrule[right]{state-step-ret}
    { \Phi(\expr, \state,  \err) }
    { \stateStep{\expr}{\state}{\err}{\Phi} }
    \and
    \infrule[right]{state-step-exp}
    {
      \expect[\mu]{\Err} \leq \err \\
      \erasable(\mu, \state_1) \\
      \All \state_2 \in \supp \mu.
      \stateStep{\expr}{\state_2}{(\Err(\state_2))}{\Phi}
    }
    { \stateStep{\expr}{\state_1}{\err}{\Phi} }
        \and
    \infrule[right]{state-step-promote}
    { \All f \in \supp (\urnsf~\state).\spac \urnsubst~f~\val= \valB
      \\
      \stateStep{(\lctx[\valB])}{\state}{\err}{\Phi} }
    { \stateStep{(\lctx[\val])}{\state}{\err}{\Phi} }
  \end{mathpar}
  \caption{Inductive Definition of the State Step Precondition $\stateStep{\expr}{\state}{\err}{\Phi}$.}
  \label{fig:stateStep}
\end{figure*}

\paragraph{Program Step Precondition}
Unlike the state step precondition previously, the program step precondition is not a fixpoint but a predicate of a single inference rule \ruleref{prog-step-exp}. Morally speaking, this rule is similar to \ruleref{state-step-exp}, for example error credits can be redistributed across some distribution according to function $\Err$. However, instead of taking an erasable step on the state, \ruleref{prog-step-exp} takes an actual execution step on the program configuration, which we use to prove the soundness of computational rules, e.g.~\ruleref{ht-pure} and \ruleref{ht-rand}.
\begin{mathpar}
  \inferH{prog-step-exp}
  { \red(\expr_1, \state_1) \\
    \expect[\stepdistr (\expr_1, \state_1)]{\Err} \leq \err \\
     \All (\expr_2, \sigma_2) \in \supp(\stepdistr (\expr_1, \state_1)).
     \Phi (\expr_2, \state_2, \Err(\expr_2, \state_2)) }
  {\progStep{\expr_1}{\state_1}{\err}{\Phi}}
\end{mathpar}
\paragraph{Promote Update Predicate}
Recall that one needs to provide a promote update predicate before promoting values (see \cref{sec:elton-solution} and \cref{sec:urn-rules}).
Under the hood, the definition of the promote update predicate mirrors elements from \ruleref{state-step-promote}.
Concretely, we can prove $\rupd{\val}{\prop}{\valB}$ if given the state interpretation, we can establish $\prop$ and the state interpretation, and prove that the urn substitution of $\val$ equals $\valB$ under all functions in the support of $\urnsf~\state$.
\begin{align*}
  \rupd{\val}{\prop}{\valB} \eqdef{}
  &\All \state, \err. 
    \stateinterp~\state~\err~ \wand (\All f \in \supp (\urnsf~\state) .\spac \urnsubst~f~\val = \valB) \sep\prop \sep \stateinterp~\state~\err
\end{align*}
\subsection{Soundness}

\label{sec:soundness}

Recall that the adequacy theorem of \theaplog (\cref{thm:adequacy} in \cref{sec:logic}) asserts that delayed sampling and urn resources  are sound extensions to the language and logic. To prove the theorem, we first prove the  following \cref{lem:commuting}.

\begin{lemma}[Commuting Lemma] \label[lemma]{lem:commuting}
  If there exists configuration $\cfg \in \supp(\stepdistr(\expr, \state))$, then
  \begin{align*}
    &
      \begin{array}{l}
        \hcode{\hcode{\urnsf~\state\mbindi}} \hcode{\Lam f. \urnsubstcfg~f~(\expr, \state) \mbindi} \\
        \spac\spac \bluecode{\Lam \cfg'. \stepdistr~\cfg'}
        \end{array}=
      \begin{array}{l}
        \bluecode{\stepdistr(\expr,\state) \mbindi} \\
        \spac\hcode{\Lam (\expr',\state'). \urnsf~\state' \mbindi}\hcode{\Lam f. \urnsubstcfg~f~(\expr',\state')}
        \end{array}
  \end{align*}
\end{lemma}
The statement of the commuting lemma concerns two separate actions: the \bluecode{blue} action represents a single step of program execution, while the \hcode{orange} action represents the replacement of delayed symbolic values with their actual values according to the substitution function $f$ induced by the distribution $\urnsf$ on the state. \cref{lem:commuting} states that these two actions commute with each other, and it does not matter which action we perform first. 

To prove the adequacy theorem of \theaplog, we first prove the following more
general result (\cref{thm:adequacy-intermediate}). We prove this result 
by induction on $n$, repeatedly unfolding the definition of the weakest
precondition and applying the \cref{lem:commuting} to swap one \bluecode{program
  execution step} with the \hcode{orange} action. \cref{thm:adequacy} can then
be recovered by only taking into account expressions $\expr$ without symbolic
delayed values (the \hcode{orange} part trivially becomes
$\mret (\expr,\state)$) and taking the limit of $n$.

\begin{lemma}\label[lemma]{thm:adequacy-intermediate}
  If $\upto{\err}\vdash \wpre{\expr}{\issimpleval~v\sep\pprop}$,
  then for all states $\state$, and natural numbers $n$,
  $\pr[\distr]{\neg \pprop} \leq \err$ where
    $\distr\eqdef{}
                   \hcode{ \urnsf~\state \mbindi \Lam f. \urnsubstcfg~f~(\expr,\state) \mbindi }
    \bluecode{\Lam \cfg. \execVal_{n}~\cfg}$.
    
\end{lemma}

\section{Related Work}
\label{sec:related-work}

\paragraph{Separation Logic for Probabilistic Programs}

In recent years there have been a number of works that use separation logic to
reason about probabilistic programs. As described in \Cref{sec:introduction}, we can distinguish these into two main classes. The first class are \emph{lifting-based logics} that extend standard separation logics
 with special types of resources to capture probabilistic
aspects of program execution, but ultimately keep the same assertion model as
the base logic. Logics in this class include 
Clutch~\cite{clutch}, Eris~\cite{eris}, Approxis~\cite{approxis},
Coneris~\cite{coneris}, and Foxtrot~\cite{foxtrot}.
While \theaplog is a lifting-based logic, urn resources make it possible to state a form of distributional assertions over delayed values.

Another line of work instead takes the approach of reinterpreting the
separating conjunction to represent probabilistic independence. These generally
consider assertions over distributions, so they can be classified as
distribution-based logics. After PSL~\cite{psl}, which pioneered the concept of
separation as independence, different extensions were developed,
e.g.~LINA~\cite{lina} or Lilac~\cite{lilac}. Bluebell~\cite{bluebell}
reintroduces some of the unary and relational lifting-based
reasoning into a distribution-based logic through the so-called joint
conditioning modality. The recent Amaryllis~\cite{amaryllis} extends this to support dynamic allocation. In addition to reasoning
about the distribution of variables, probabilistic
variants of Outcome Logic~\cite{dol, pcol} support reasoning about nondeterminism and concurrency. As
discussed before, while these logics can naturally express distributional
invariants, they often consider simplified languages (e.g. without unbounded
loops, local state, or higher-order functions) and their assertion models are
not rich enough to express adversarial reasoning principles, which makes
verifying our case studies out of their scope.

\paragraph{Prophecy Variables and Presampling Tapes}

One of the key challenges that \theaplog tackles is reasoning about the result
of a probabilistic sampling when the concrete instantiation of the program
logic rule (\ruleref{ht-rand}) depends on some event that happens later in the execution, in this
case, an adversary call. The need to guess future nondeterministic behaviors
motivated the development of \emph{prophecy variables}~\cite{prophecy1} which
were later introduced in Iris~\cite{iris-prophecy}. A related motivation
in the setting of probabilistic programs led to the introduction of
\emph{presampling tapes} in Clutch~\cite{clutch}. Presampling tapes can be used
to generate the randomness in advance, that is used later in a program.
While Eris supports presampling, this is not enough to reason about the
classes of examples we consider; as discussed in \cref{sec:eris-issue},
we want to delay the resolution of random samplings, not advance it.

\paragraph{Reasoning about Unknown Code}

There is a vast literature on robust safety and reasoning in the presence of
unknown code~\cite{robust-safety-gordon, robust-safety, cerise, secref}. 
Here we focus on some instances of adversarial reasoning principles in
probabilistic program logics. Adversary rules are available in some logics such
as apRHL~\cite{aprhl2}, aHL~\cite{union-bound-logic} or Ellora~\cite{ellora},
often with security applications in mind. These logics consider a simple model
of global first-order state, and rules enforce some ad-hoc syntactic
restrictions on the adversary to ensure that private fragments of memory are
not shared. \citet{HO-UBL} introduce adversary rules in a setting with first-order
global state where adversary separation is enforced through a graded type
system.

In general, adversarial reasoning principles can be derived in any logic with a
logical relations model, since the soundness theorem derives a
specification for any well-typed program. For example Approxis~\cite{approxis} and
Foxtrot~\cite{foxtrot} have logical relations models to reason about contextual
equivalence. However, they lack the expressivity to capture the kind of
probabilistic invariants we study.

SSprove~\cite{ssprove} is a mechanization of state-separating
proofs~\cite{DBLP:conf/asiacrypt/BrzuskaDFKK18}, a technique to reason about
security properties. The idea is to encapsulate different components of the
system in packages, with interfaces to communicate with each other. Adversaries
can then be implemented as packages themselves, with a state that is disjoint
from the rest of the system, and their behavior can be restricted through the
interface that is exposed to them. It considers a monadic language, which is
not expressive enough to cover the kinds of features we consider.

\section{Conclusion}
\label{sec:conclusion}
We presented the higher-order separation logic \theaplog,  based around two novel extensions to previous work: a modified
operational semantics with delayed samplings at the language level, plus the
corresponding urn resources and rules at the logic level. We connected  these extensions back to standard
 operational semantics through the adequacy theorem, and showed
that they are enough to capture many properties of interest that are out of
scope of previous work.

On the other hand, the approach is arguably convoluted
and it is not easy to see how it would scale for additional probabilistic features.
One limitation of this approach so far is that urns are restricted to uniform distributions.
It would be interesting to support more
flexible urns that describe more
complicated distributions, e.g.~those that are non-uniform or countably infinite.
We would also want to support some form of urn mixing where we combine
urns arising from separate delayed sampling instructions.
We believe that these extensions require significant modifications to the model
of \theaplog, where one would need to formalize \theaplog with more foundational measure theory.
Nonetheless, as we showed with a wide range of case studies,
 we  see \theaplog as a promising approach that shows how to express
distributional properties inside a lifting-based logic.

\begin{acks}
  This work was supported in part by the \grantsponsor{NSF}{National Science Foundation}{}, grant no.~\grantnum{NSF}{2338317}, \grantsponsor{Villum}{Villum}{} Investigator grants, no. \grantnum{Villum}{VIL25804} and no. \grantnum{Villum}{VIL73403}, Center for Basic Research in Program Verification (CPV), from the VILLUM Foundation, and the European Union (\grantsponsor{ERC}{ERC}{}, CHORDS, \grantnum{ERC}{101096090}).
  Views and opinions expressed are however those of the author(s) only and do not necessarily reflect those of the European Union or the European Research Council.
  Neither the European Union nor the granting authority can be held responsible for them.
\end{acks}

\bibliography{refs}

\ifbool{fullversion}{
  \pagebreak
  \appendix
  \section{Counter Example of Prophecy Variables}
\label{appsec:prophecy}
We show that the standard rules of prophecy variables in the context of Iris~\cite{irisjournal}
are unsound in the unary setting of Elton.
The rules applied in this proof are only from the Eris~\cite{eris} fragment of the logic, and hence
is unsound already in the Eris logic.

\newcommand{\NewProph}{\operatorname{\langkw{NewProph}}}
\newcommand{\ResolveProph}[2]{\operatorname{\langkw{Resolve}} #1 \operatorname{\langkw{to}} #2}
\newcommand{\Proph}{\operatorname{\mathsf{Proph}}}
\newcommand{\counterprog}{\mathit{proph\_rand}}

Assume the existence of two operators $\NewProph$ and $\ResolveProph{p}{b}$ in our programming language and their (unsound) Hoare-triple specifications found below.
\begin{mathpar}
  \inferH{wp-newproph-unsound}
  {}
  { \hoare{\TRUE}{\NewProph}{ p . \exists n \in \nat . \Proph(p, n) }[\mask]}

  \inferH{wp-resolve-unsound}
  {}
  { \hoare{ \Proph(p, n) \ast n' \in \nat }{\ResolveProph{p}{n'}}{b = n'}[\mask] }
\end{mathpar}
These specifications capture \emph{one-shot prophecies} for natural numbers. The first rule \ruleref{wp-newproph-unsound}
generates a fresh prophecy variable and a prophecy resource $\Proph(p,n)$. When we subsequently
execute $\ResolveProph{p}{n'}$, \ruleref{wp-resolve-unsound} can then resolve the prophecy variable to the value $n'$
and assert $n=n'$. Now consider the following program: 
\begin{align*}
  \counterprog \eqdef{}
  &\Let p = \NewProph in \\
  &\Let n = \Rand 1 in \\
  &\hspace{-0.15em}\ResolveProph{p}{n}
\end{align*}
This program $\counterprog$ first creates a new prophecy and samples a number $n$ uniformly between $0$ and $1$. It then finally resolves the prophecy variable to $n$. With the above rules for prophecy variables, we can prove the following Hoare triple, stating that whenever the program terminates, we can prove $\FALSE$, with error probability at most $\frac{1}{2}$. This is a contradiction since $\counterprog$ terminates safely with probability $1$!
\begin{mathpar}
  \hoare{\upto{\frac{1}{2}}}{\counterprog}{\FALSE}
\end{mathpar}
The proof of this Hoare triple is relatively straightforward. After applying \ruleref{wp-newproph-unsound} to
create a fresh prophecy variable prophesying number $n$, we apply \ruleref{ht-rand-avoid} to avoid sampling that particular $n$ with the provided error credit. Finally, with \ruleref{wp-resolve-unsound} asserting that the sampled value is the same as $n$, we derive a contradiction.

  \section{Proof of Preimage Resistance of Random Oracle}
\label{appsec:preimage-resistance-proof}
We present the invariant for the proof of the preimage resistance of the random oracle example in \cref{sec:preimage-resistance}.
\begin{align}
  I~\gname\eqdef& \Exists (i':\tnat)~(m:\tint \fpfn \tint). \nonumber \\
                & \quad \hashfun~\hashf~(\mapinsert{\ulbl}{h}{m}) \sep \label{eq:hashprog-inv-1}\\
                & \quad \progheap{\loc}{i'} \sep i'\leq \tapeboundB \sep \label{eq:hashprog-inv-2}\\
                & \quad \Exists \urn. \textlog{disjoint}~\urn~(\dom~m) \sep \urn\neq\emptyset\sep \progurn{\ulbl}{\urn} \sep \label{eq:hashprog-inv-3}\\
                & \quad \left( \left(
                  \begin{array}{l}
                    (\All x. x\in\cod (m) \ra x\neq h) \sep \\
			  \upto{\frac{i'+1}{\maxsize+1}}\sep\upto{\frac{i'+1}{\ssize{\urn}}} \sep\\
			  \tapebound+1+i'-\tapeboundB\leq\ssize{\urn}
                  \end{array}
                  \right)\lor \ownGhost{\gname}{\token} \right)\label{eq:hashprog-inv-4}
\end{align}
Let us go through each line of the invariant. Starting off with \eqref{eq:hashprog-inv-1}, the invariant stores a $\hashfun$ predicate, asserting that $\hashf$ can be represented by some map $m$ with $h$ inserted under the urn label $\ulbl$. Notice that $m$ is a map from integers to integers, as it describes the queries called by the adversary which can only be queries of integers as described by its type, but not any delayed value. The second line \eqref{eq:hashprog-inv-2} describes the reference in the wrapper of $\hashf$, storing a reference resource and asserting an inequality stating that the natural number stored by the reference is smaller than the provided limit $\tapeboundB$. The next line \eqref{eq:hashprog-inv-3} describes the urn resource which stores the current potential values of the secret generated by $\DRand$. On top of storing the urn resource, we also enforce that the domain of $m$ is disjoint from $\urn$, which is reasonable as we do want our secret to be disjoint from all the guess attempts of the adversary. The final line \eqref{eq:hashprog-inv-4} is reminiscent of the disjunction in the invariant for the proof of $\cflip$, where we store a token on the right-hand side of the disjunction. On the left, we store various useful resources for bookkeeping and resolving urns. For example, we have a condition asserting that every value stored in $m$ is not equal to $h$, preventing a collision with the hash of $\secret$. Moreover, it contains some error resources and an inequality ensuring that the size of $\urn$ is not too small, so we still have enough ``space'' in the urn to execute the remaining resolve actions.

Once the exclusive token and invariant is set up, the waltz of the proof dances out smoothly to the tune of previously-seen examples. Compared to previous examples, the only part that is different is that concerning the hash function. Whenever we call $\hashf$, whether it be one of the $\tapeboundB$ calls by the adversary or the final one at the end of $\hashprog$, we do a case distinction on whether the key $k$ is called before.  In the case where it is not, we apply \ruleref{ht-hash-new} and pay $\upto{\frac{1}{\maxsize}}$ credits to ensure that the result produced is not equal to $h$. In the other case where $k$ can be resolved to some key queried before, we do another case distinction on whether $k$ can be resolved to be equal to some value in our urn $\urn$. If it is, we apply \ruleref{ht-resolve-partial-avoid} to resolve the urn to not contain $k$, splitting error credits accordingly just as how it was done with $\interactiveguess$ in \cref{sec:interactive-guessing}. Otherwise, this means that $k$ is some key that is queried before by the adversary, and we can call \ruleref{ht-hash-prev} to return the value stored under $k$, which is not equal to $h$ by the collision-resistance condition in the invariant \eqref{eq:hashprog-inv-4}.

  \section{Invariant for Proof of Unlinkability of Basic-Hash}
\label{appsec:basic-hash-proof}
In this section, we state the invariant needed to prove the unlinkability property of the basic-hash protocol outlined in \cref{sec:basic-hash}.

Looking at the invariant below, we observe similarities between this invariant and the one in \cref{appsec:preimage-resistance-proof}.
For example, we have the $\hashfun$ predicate tracking the contents of the random oracle at \eqref{eq:unlink-inv-1} and the reference counter tracking the number of queries made by the adversary at \eqref{eq:unlink-inv-2}. We call attention to the two special values $\val_1$ and $\val_2$ inserted into the map, representing the delayed values of $h_1$ and $h_2$, respectively.
The conditions \eqref{eq:unlink-inv-3}, \eqref{eq:unlink-inv-4}, and \eqref{eq:unlink-inv-5} describe relevant information about urn resources, e.g.~the size of the urns is bounded below, as well as the error credits for their resolution. Lastly, we have the familiar-looking disjunction in \eqref{eq:unlink-inv-6}. On the left disjunct, we have the unresolved urn for $b$ and $\upto{\frac{1}{2}}$, while on the right disjunct, we resolve the urn to a singleton value by spending the error credit to avoid the value returned by the adversary with \ruleref{ht-resolve-all-avoid}. 
\begin{align}
  \val_1\eqdef{}&n_1 \cdot (\tapebound_{T}+1) +_{\delayv} \ulbl \notag\\
  \val_2\eqdef{}&n_2 \cdot (\tapebound_{T}+1)+_{\delayv}((b \cdot_{\delayv} \ulbl) +_{\delayv} ((1-_{\delayv} b) \cdot_{\delayv} \ulbl'))\notag \\
  \notag\\
  I~\gname\eqdef{}& \Exists (i':\tnat)~(m:\tint \fpfn \tint)~x~y. \nonumber \\
                & \quad \hashfun~\hashf~(\mapinsert{\val_2}{x}{\mapinsert{\val_{1}}{y}{m}}) \sep \label{eq:unlink-inv-1}\\
                & \quad \progheap{\loc}{i'} \sep i'\leq \tapeboundB \sep \label{eq:unlink-inv-2}\\
                & \quad \Exists \urn. \left(\All z. z \in \urn \ra
                  \left(
                  \begin{array}{l}
                    n_1\cdot (\tagrange+1) + z \notin \dom(m) \land \\
                    n_2\cdot (\tagrange+1) + z \notin \dom(m) \land \\
                    0\leq z \leq \tagrange
                  \end{array}
                  \right)
                  \right) \sep \label{eq:unlink-inv-3}\\
                & \quad s\neq \emptyset \sep \progurn{\ulbl}{\urn} \sep \progurn{\ulbl'}{\urn} \sep \label{eq:unlink-inv-4} \\ 
		& \quad \upto{\frac{2\cdot i'}{\ssize{\urn}}}\sep (\tagrange+1 - (\tapeboundB-i')\leq \ssize{\urn}) \sep \label{eq:unlink-inv-5} \\ 
                & \quad \left( \left(
                  \begin{array}{l}
                    \progurn{b} {\{0, 1\}} \sep\\
                    \upto{\frac{1}{2}} 
                  \end{array}
                  \right)
                  \lor
                  \left(
                  \begin{array}{l}
                    \ownGhost{\gname}{\token} \sep \\
                    \Exists (k:\tnat). k\leq 1 \sep \\
                    \progurn{b}{\singleton{k}}
                  \end{array}
                    \right) \right)\label{eq:unlink-inv-6}
\end{align}

  \section{Proof of Generic Group Model}
\label{appsec:generic-group-proof}
In this section, we present details of the generic group model example, briefly mentioned in \cref{sec:generic-group}.

We start by implementing a model of the generic group model using a mutable map, see \cref{fig:generic-group-code}. Internally, we represent group elements by their exponent in the mutable map and identify them by their indexes in the map. The function $\arrget$ reads from the mutable map when provided an identifier and returns the exponent of the group element identified. To create a new identifier for a potentially new group element, we extend the map with a fresh identifier and return it. Note that it is possible to have a group element identified by two distinct identifiers, but this does not pose an issue with distinguishing them; as we later see, we apply $\veq$ to compare whether two identifier represent the same group element.

To bound the computational power of the adversary, we deploy a similar trick as that in \cref{sec:preimage-resistance} and \cref{sec:basic-hash} where we use a reference to upper bound the number of times it can perform various group operations. We define the function $\vtryspend$ that returns $\True$ if we can decrement a reference, and return $\False$ if the reference is already non-positive.

For this example, the generic group model exposes two main group operations. We can multiply two group elements with $\vmul$, which reads two values from map and stores the sum of the values modulo $p$. We can also take the inverse of a group element with $\vinv$ by reading a value from the map and storing the subtraction of it from $p$ modulo $p$. In our example, we  pass $\vtryspend~\loc$ for some reference $\loc$ as the argument $f$ in the functions, thus upper bounding the number of times a generic algorithm (the adversary) can perform these operations. Note also that the functions crash with $\alertcode{\TT~\TT}$ in the case we provide an identifier not in the domain of the mutable map; in theory, this branch is never executed because all identifiers returned by the functions are valid and the adversary has no ability to modify or create new identifier other than the identifiers returned by functions exposed by the generic group model.

Lastly, the model also exposes a $\veq$ function that checks whether two identifiers match to the same group element according to the mutable store. Unlike $\vmul$ or $\vinv$, we do not bound the number of times the adversary can call $\veq$ as we argue an adversary in the real world should be able to distinguish group elements syntactically without consuming significant computational power.

\begin{figure}[ht]
  \begin{subfigure}[t]{0.43\textwidth}
    \begin{align*}
      \arrnew~\_\eqdef{}& \Alloc (0, \initmap~\TT)\\
        \arrpush~\varr~\val \eqdef{}& \Let (i, lm) = \deref \varr in \\
                                   & \mapset\ lm\ i\ \val;    \\
                                   & \varr\gets (i+1, lm); i
    \end{align*}
  \end{subfigure}
  \begin{subfigure}[t]{0.43\textwidth}
    \begin{align*}
        \arrget~\varr~i \eqdef{}& \Let (\_, lm) = \deref \varr in \\
                                   & \mapget\ lm\ i    \\
        \vtryspend~\loc~\_ \eqdef{}& \Let i = \deref \loc in \\
                                   & \If i\leq 0 then \False \\
                                   & \quad \Else \loc\gets (i-1); \True
    \end{align*}
  \end{subfigure}
  \begin{subfigure}[t]{0.43\textwidth}
    \begin{align*}
      \vmul~\varr~f~i_1~i_2\eqdef{}& \If f~\TT then \\
                           & \qquad \Let e_1 =\arrget~\varr~i_1 in\\
                                   & \qquad \Let e_2 =\arrget~\varr~i_2 in\\
                                   & \qquad \MatchML (e_1, e_2) with | (\Some a, \Some b) => \Some (\arrpush~\varr~((a+b)~\textlog{mod}~p)) | \_ => \alertcode{\TT~\TT} end{}\\
                           & \quad \Else \None\\
      \vinv~\varr~f~i\eqdef{}& \If f~\TT then \\
                           & \qquad \Let e =\arrget~\varr~i in\\
                                   & \qquad \MatchML e with | \Some a => \Some (\arrpush~\varr~((p-a)~\textlog{mod}~p)) | \_ => \alertcode{\TT~\TT} end{}\\
                           & \quad \Else \None\\
      \veq~\varr~i_1~i_2\eqdef{}&  \Let e_1 =\arrget~\varr~i_1 in\\
                                   & \Let e_2 =\arrget~\varr~i_2 in\\
                                   &  \MatchML (e_1, e_2) with | (\Some a, \Some b) =>   (a=b) | \_ => \alertcode{\TT~\TT} end{}
    \end{align*}
  \end{subfigure}
  \caption{Code of generic group model. }
  \label{fig:generic-group-code}
\end{figure}

With the ingredients of the generic group prepared at our disposal, we recall the implementation of the $\dlog$ game presented in \cref{sec:generic-group} (see \cref{fig:discrete-log-code}). After taking in an adversary $\adv$, the $\dlog$ program begins by picking a random group element exponent $x$ via $\Rand (p-1)$ (modified to be $\DRand (p-1)$ for the verification part). We then initialize the generic group oracle model and add $1$ and $x$ into the oracle to get the identifier for the generator and our secret group element, respectively. We also allocate a reference $\Alloc~\tapeboundB$ which is passed to $\vtryspend$ to create a closure for bounding the number of group operations the adversary can perform. We then wrap the identifiers and the group functions into an existential value to be passed to the adversary. At last, we compare the return value of the adversary, its guess, with our original $x$.

In this example, we require that the adversary $\adv$ is well-typed with the existential type $(\Exists \type. \type\times\type\times(\type\ra\type\ra\toption~\type)\times(\type\ra\toption~\type)\times(\type\ra\type\ra\tbool))\ra\tint$. Here the existential type is crucial in capturing the idea that the only operations a generic algorithm can perform are those with the oracle, i.e.~$\vmulf$, $\vinvf$, and $\veqf$. If we instead give $\tint$ as the type of the identifiers, an adversary can modify or create an invalid identifier, which potentially leads us to execute the invalid unsafe branches containing $\alertcode{\TT~\TT}$ that are theoretically unreachable. We believe that this is a strong feature of \theaplog that its higher-order assertion logic is expressive enough to describe the logical relations of existential types, which is a second-order proposition.

Since the discrete logarithm problem is assumed to be difficult, it seems natural that we want to prove that the program returns $\False$ with negligible error probability, but what is a reasonable upper bound for that error probability? To answer this question, we must figure out at a high level where and how error credits are used in the example. The key insight boils down to what $\veqf$ should return when provided with two arbitrary identifiers. Obviously, if the two (delayed) values stored represent the same abstract number, e.g.~$1+_v \ulbl$ and $\ulbl +_v 1$, we should return $\True$. What about values that represent different abstract numbers? Consider the scenario where $p=5$ and we want to compare $1 +_v \ulbl$ and $2 +_v (2 \times_v \ulbl)$. If $\veqf$ returns $\True$ for this comparison, the adversary can immediately deduce that the secret can only be one single value for the equality to hold, $4$ in this case. At a high level, we want to use our error credits to ensure that the results of $\veqf$ reveals as little information as possible.  More precisely, if the two (delayed) values represent the same number under all possible values of the secret, which we call a \emph{trivial collision}, we return $\True$. Otherwise, we pay some error credit to shrink the urn (by at most one number) and prevent a \emph{non-trivial collision}, i.e.~to ensure that the values do not represent the same number after promotion of the urn resource.

We  assume a worst case scenario where we want to avoid non-trivial collisions for all pairs of elements inserted into the map. For each pair, there is exactly one value of the secret that we want to avoid, as it may induce a non-trivial collision. One must also not neglect the final error credit needed to account for the final guess $x'$ by the adversary. This gives us the following Hoare triple, which matches the theoretical $\mathcal{O}(\tapeboundB^2/p)$ error bound complexity. 
\begin{mathpar}
  \hoare{\upto{\frac{2+\tapeboundB\cdot(\tapeboundB+3)/2}{p}}}{\dlog~\adv}{v. v=\False}
\end{mathpar}
Before diving into  details of the proof, let us answer a question regarding the existential type of the adversary: what is the proposition we should instantiate to the abstract type $\type$ of identifiers such that we can ensure all identifiers produced map to valid values in the generic group model? To answer this, we introduce another kind of ghost resource, known as  $\authm(\nat_\text{max})$, whose rules we show in $\cref{fig:auth-frag-rules}$. The resources of  $\authm(\nat_\text{max})$ is split into two components, the authoritative part $\authfull n$  and the fragmental part $\authfrag n$. Intuitively speaking, the authoritative part represents a monotonically-increasing counter, while the fragment represents knowledge of the lower bound of the counter. At any point of a proof, we can apply \ruleref{ht-auth-frag-alloc} to create the authoritative and fragmental part with the same value $n$. We can also increase the value stored of an authoritative part through \ruleref{ht-auth-update}. The rule \ruleref{auth-frag-valid} captures the idea that the fragmental part is always a lower bound on the value in the authoritative part. Lastly, the fragmental part is duplicable (\ruleref{frag-dup}), for in the idealistic but unrealistic Utopian world of Iris, knowledge is free to use, reuse, and redistribute. In our proof, we later store the authoritative part in our invariant to describe the size of the mutable map, and place the fragmental part in the interpretation of the abstract identifier type. Specifically, we define the value interpretation of the existential variable in the adversary type to be $\semInterpS{\type}{\Delta}\eqdef{} \Lam v. \Exists n\in\nat.v=n \sep \authfrag n $. By \ruleref{auth-frag-valid}, we can show any identifier returned by $\vmulf$ and $\vinvf$ are in the domain of the generic group map. 

\begin{figure}[ht]
  \centering
\begin{mathpar}
  \inferH{ht-auth-frag-alloc}
  {\hoare{\authfull n \sep \authfrag n \sep \prop}{\expr}{\propB}[\mask]}
  {\hoare{\prop}{\expr}{\propB}[\mask]}\and
  \inferH{ht-auth-update}
  {n\leq n' \\ \hoare{\authfull{n'}\sep\authfrag{n'}\sep\prop}{\expr}{\propB}[\mask]}
  { \hoare{\authfull{n}\sep\prop}{\expr}{\propB}[\mask]}
  \and
  \inferH{auth-frag-valid}
  {}
  {\authfull m \sep\authfrag n \vdash n\leq m}
  \and 
  \inferH{frag-dup}
  {}
  { \authfrag n \vdash \authfrag n \sep \authfrag n}
\end{mathpar}
  \caption{Selected rules of $\authm(\nat_\text{max})$. }
  \label{fig:auth-frag-rules}
\end{figure}

Albeit more complicated, the invariant for the generic group model program follows a similar structure as that for previous examples (e.g.~\cref{sec:preimage-resistance,sec:basic-hash}), in particular we observe the distinctive disjunction in \eqref{eq:generic-group-inv-5}, used by the main program to swap its token resource for various other  resources for urn resolution. On the left disjunct, we state that there are no non-trivial collisions in the abstract map $m'$ via $\nocoll$ and that the size of our urn is not too small, i.e.~there are still many possibilities of what our secret group element can be. The left disjunct also contains error credits to ensure that the map remains non-trivial collision-free as the adversary inserts more elements into the map via the remaining group operations.

Let us look at the other lines of the invariant. The first three lines of the invariants are relatively unsurprising. We track the urn resource for our secret group element in \eqref{eq:generic-group-inv-1}, relate the size of the mutable map and with the value stored in $\loc$ in \eqref{eq:generic-group-inv-2}, and represent the mutable map $lm$ with an abstract map $m$ in \eqref{eq:generic-group-inv-3}. The next line \eqref{eq:generic-group-inv-4} involves the use of another map $m'$ which  has the same domain as $m$, and is used to keep track of the number represented by the delayed values stored in $m$. Specifically, the predicate $\mapmatch$ asserts that that number stored in $m$ can be abstractly represented by a unique linear combination of $\ulbl$ and $1$, with coefficients $a$ and $b$ between $0$ and $p-1$ inclusively, respectively, i.e.~$a\cdot \ulbl + b$. Additionally, we store the authoritative resource, that ensures all generated identifiers are in the domain of our map. 
\begin{align}
  I~\gname\eqdef& \Exists (m:\tint \fpfn \val)~(m':\tint \fpfn (\tnat\times\tnat))~(i':\tnat)~\urn. \nonumber \\
                & \quad \progurn{\ulbl}{\urn}\sep \urn\neq\emptyset\sep(\All x. x\in\urn \ra 0\leq x<p)\sep \label{eq:generic-group-inv-1}\\
                & \quad \progheap{\varr}{(i,lm)}\sep (2\leq i'\leq \tapeboundB+2) \sep \progheap{\loc}{(\tapeboundB+2-i')}\sep \label{eq:generic-group-inv-2}\\
                & \quad \mapfun~lm~m \sep \dom(m)=\{0,\dots,i'-1\} \sep (\All \val. \val\in\cod(m)\ra \typef(\val)=\tint) \sep\label{eq:generic-group-inv-3}\\
                & \quad \dom(m)=\dom(m')\sep (\All a~b.(a,b)\in \cod(m')\ra 0\leq a <p \land 0\leq b < p) \sep \label{eq:generic-group-inv-3}\\
                &\quad \mapmatch~m~m'~\ulbl\sep \authfull~(i'-1) \sep \label{eq:generic-group-inv-4}\\
                & \quad \left( \left(
                  \begin{array}{l}
                    \nocoll~m'~\urn\sep\\
			  (p\leq \ssize{\urn} + \frac{i' \cdot (i'+1)}{2}) \sep \\
			  \upto{\frac{1+ (\tapeboundB+2-i')\cdot(\tapeboundB+i'+1)/2}{\ssize{\urn}}}
                  \end{array}
                  \right)\lor \ownGhost{\gname}{\token} \right)\label{eq:generic-group-inv-5}
\end{align}
Once the invariant is figured out, the rest of the proof is simple, if not obvious! Each time the adversary performs a group operation to extend the map by one more element, we apply some error credits to ensure that there is no non-trivial collision between that element and every other element already inserted. In other words, the application of \ruleref{ht-resolve-partial} here is similar to that in \cref{sec:interactive-guessing} and \cref{sec:preimage-resistance}, with the slight caveat that we are not just avoiding one element, but multiple elements (every other element in the map) for each call of a group operation (which eventually amounts to applying \ruleref{ht-resolve-partial-avoid} multiple times sequentially). 

}{}

\end{document}